\documentclass[preprint,12pt]{elsarticle}

\usepackage{ulem}
\usepackage{amssymb}
\usepackage{amsmath}
\usepackage{breqn}
\usepackage[section]{placeins}
\usepackage{graphicx}
\usepackage[colorlinks=true, allcolors=blue]{hyperref}

\usepackage{caption,subcaption,tikz}
\usetikzlibrary{arrows.meta}
\DeclareCaptionFormat{suboverlay}{\gdef\subcapoverlay{(\thesubfigure) #3\par}}
\DeclareCaptionStyle{suboverlay}{format=suboverlay}

\newcommand{\subcaptionOverlay}[1]{%
	\subcaption{}%
	\begin{tikzpicture}
		\node [inner sep=0,anchor=north west]at (-2ex,2ex) (image) {#1};
		\draw node [black] {\subcapoverlay};
	\end{tikzpicture}%
}

\journal{CMAME}

\begin{document}
	
	\begin{frontmatter}
		
		
		
		\title{Optimization-based Level-Set Re-initialization: A Robust Interface Preserving Approach in Multiphase Problems}
		
		\author[inst2]{Alireza Hashemi}
		\ead{ahashemi@cimne.upc.edu}
		
		\author[inst2]{Mohammad R. Hashemi}
		\ead{mhashemi@cimne.upc.edu}
		
		\author[inst1,inst2]{Pavel B. Ryzhakov\corref{cor1}}
		\ead{pavel.ryzhakov@upc.edu}
		
		\author[inst1,inst2]{Riccardo Rossi}
		\ead{rrossi@cimne.upc.edu}

		\cortext[cor1]{Corresponding author}
		\affiliation[inst2]{organization={Centre Internacional de Mètodes Numèrics en Enginyeria (CIMNE)},
			addressline={Gran Capitan, S/N}, 
			city={Barcelona},
			postcode={08034},
			country={Spain}}
		
		\affiliation[inst1]{organization={Universitat Politècnica de Catalunya (UPC)},
			addressline={North Campus}, 
			city={Barcelona},
			postcode={08034},
			country={Spain}}
		
		\begin{abstract}
			In spite of its overall efficiency and robustness for capturing the interface in multiphase fluid dynamics simulations, the well-known shortcoming of the level-set method is associated with the lack of a systematic approach for preserving the regularity of the distance function.
			This is mainly due to the stretching (or compressing) effect of the strain rate especially in the vicinity of the liquid-gas interface.
			Level-set re-initialization is an effective treatment for this issue.
			However, the traditional approach based on the hyperbolic Hamilton-Jacobi equation is a computationally expensive procedure.
			Crucially, due to the hyperbolic nature of the formulation, the accuracy of the results hinges significantly on the specialized handling of blind spots near the liquid-gas interface intersecting the substrate.
			The present work proposes a two-step elliptic level-set re-initialization approach that strictly preserves the location of zero level-set via incorporation of an element splitting process.
			The primary initialization step helps remove any non-smoothness in the to-be regularized level-set function dramatically improving the efficiency of the secondary optimization step.
			Geometric representation of the boundary conditions is utilized in the initialization step, while the optimization step minimizes the reliance of the results on the treatment of the blind spots.
			The performance of the proposed method is examined for free and sessile three-dimensional droplets.
		\end{abstract}
		
		
		
		
	\end{frontmatter}
	
	
	\section{Introduction}
	Since its first introduction in the context of computational fluid dynamics (CFD)~\cite{mulder_computing_1992}, the level-set method~\cite{osher_fronts_1988,gibou_review_2018} has extensively been used for multi-phase flow problems~\cite{sussman_level_1994,osher_level_2001,hashemi_toward_2021,noda_extended_2022}.
	By capturing the phase interface evolution on a fixed computational mesh, the level-set method circumvents the need for complicated deformation and regeneration of the mesh. 
	While the alternative technique, volume of fluid (VOF)~\cite{hirt_volume_1981}, is usually preferred in the context of the finite volume method~\cite{maric_unstructured_2020,aniszewski_parallel_2021}, the level-set method is the most widely used technique in the context of the finite element simulation of multi-phase flows~\cite{hashemi_enriched_2020,maarouf_characteristicsfinite_2022}.
	This is mainly due to the ease of calculating the major geometrical properties of the phase interface, i.e., the normal vector and curvature.
	
	The main drawbacks of the level-set method are the lack of the mass conservation property~\cite{sussman_coupled_2000} and its proneness to irregularities~\cite{hashemi_enriched_2020}.
	The former is usually alleviated by using higher-order level-set convection approaches~\cite{solomenko_mass_2017,hashemi_enhanced_2022} and introducing corrective terms to the formulation~\cite{yap_global_2006,ge_efficient_2018}.
	On the other hand, the irregularities are mainly caused by the local expansion and contraction of the distance between the convected surfaces of iso-level-set functions~\cite{trujillo_distortion_2017}.
	This \textit{per se}, is an expected result of a non-zero strain rate corresponding to the convection velocity field.  
	In order to address such irregularities, one can either prevent (or minimize) their occurrence during the convection of the level-set~\cite{janodet_massively_2022,larios-cardenas_deep_2021}, or, alternatively, frequently apply a level-set (distance) re-initialization process~\cite{xue_new_2021}.
	While the former treatment has not yet been widely assessed, the latter option has become the common way of dealing with the irregularities in the convected level-set function. 
	It is worth mentioning that in order to minimize the aforementioned drawbacks, some works have also been dedicated to the incorporation of the (semi-)Lagrangian interface tracking approaches, e.g., the particle level-set method~\cite{gaudlitz_improving_2008}.
	
	The preservation of the configuration of the zero level-set surface is of utmost importance for a robust level-set re-initialization method~\cite{sussman_efficient_1999}, as it defines the phase interface in the context of two-phase flow. Failure to maintain the zero level-set during distance re-initialization not only introduces geometrical inaccuracies but also undermines mass conservation~\cite{solomenko_mass_2017,zhang_efficient_2020}.
	Therefore, the so-called ``anchoring'' of the interface~\cite{ramanuj_high_2018} is of essential importance in the development of distance re-initialization techniques.
	
	The conventional approach used level-set re-initialization approach relies on the iterative solution of the Eikonal equation~\cite{sussman_level_1994,sethian_fast_2000,karakus_local_2022}, characterized by a hyperbolic mathematical formulation governing the regular propagation of the iso-level-set surfaces~\cite{hysing_new_2006}. While widely used, this method exhibits poor efficiency when the initial state deviates significantly from the signed distance function~\cite{min_reinitializing_2010}.
	Moreover, for severely irregular cases, the preservation of the configuration of the zero level-set cannot be guaranteed ~\cite{li_level_2005}.
	A potential remedy to the above-mentioned issue would be to rely on the geometrically evaluated distance to the fixed zero level-set.~\cite{henri_geometrical_2022}.
	Nevertheless, these approaches that basically reproduce the minimum distance would suffer from blind spots in the vicinity of the cut in the zero level-set, leading to failure in scenarios such as droplet dynamic simulations involving a contact line~\cite{della_rocca_level_2014}.
	In an alternative approach, Elias et al.~\cite{elias_simple_2007} proposed to march over the elements (computational cells) and analytically calculate the unknown nodal value of the level-set function.
	This process that starts from the vicinity of the interface (zero level-set), however, occasionally disturbs the smoothness of the level-set function.
	Furthermore, sweeping the whole domain, far from the interface, makes the procedure computationally expensive.
	
	Another class of potentially effective but infrequently utilized level-set re-initialization methods is based on solving an optimization problem~\cite{li_distance_2010,basting_minimization-based_2013,basting_optimal_2017}.
	Governed by an elliptic equation, these methods do not encounter blind spots in the domain.
	By employing such a method, preserving the configuration of the zero level-set can be achieved through the introduction of appropriate penalty terms.
	Nevertheless, the non-linearity of the governing equation increases the sensitivity of the optimization procedure to the initial state (of the level-set function).
	Improper treatment of a severely non-regular initial state can result in high computational costs or even lead to the failure of the optimization procedure.
	
	To address the limitations of optimization-based level-set re-initialization approaches, this work proposes a novel strategy, which begins by employing a reconstruction scheme, ensuring a smooth initial state for the subsequent optimization process. In essence, our approach consists of an elliptic initialization in the first step, followed by optimization of the initial guess in the second step.
	A geometric representation of the level-set gradient featuring a sophisticated blind spot treatment is proposed to handle the Newmann boundary condition required by the primary elliptic re-initialization step.
	The optimization process, \textit{per se}, is also enhanced by introducing a modified potential function.
	Moreover, the proposed method utilizes an element splitting procedure, which facilitates the accurate incorporation of the interfacial penalty terms and consequently, ensures the preservation of the zero level-set configuration during the re-initialization process.
	
	In the following section, the optimization process along with the proposed potential function are elaborated.
	Then, the elliptic reconstruction scheme and the imposition of the boundary conditions are detailed, focusing on the geometrical approach proposed to deal with the blind spots.
	Before closing this section, the element splitting technique is also briefly described.
	The method is verified through a series of static droplet simulations, and finally, its performance is shown for some dynamic cases. 
	
	\section{Formulation}
	Since its first application in the context of the multi-phase methods~\cite{sussman_level_1994,osher_level_2001}, the level-set method has been known as a competent phase-interface capturing approach~\cite{gibou_review_2018}.
	Its application is based on the convection of smooth field variable $\phi(\mathbf{x})$ representing the signed distance to the interface ($\left|\nabla\phi\right|=1$).
	This reads
	\begin{equation}\label{eq:level-set}
		\frac{\partial \phi}{\partial t}+\mathbf{u}\cdot\nabla{\phi} = 0,
	\end{equation}
	where $\mathbf{u}(\mathbf{x})$ is the associated velocity field.
	Here, $\mathbf{x}$ is the location vector within the problem domain of $\Omega \subset \mathbb{R}^d$ with $d$ denoting the spatial dimensions.
	The level-set method works well while the regularity of the level-set field is preserved.
	In most cases, $\mathbf{u}$ deviates from a uniform velocity field characterized by a zero strain rate, i.e., $\nabla\mathbf{u}=0$).
	This explains the expansion and contraction of the distance between iso-$\phi$ surfaces in the area with positive and negative strain rates, respectively.
	In the mathematical terms, $\mathbf{n}\cdot\left(\mathbf{n}\cdot\nabla\mathbf{u}\right)>0$ and $\mathbf{n}\cdot\left(\mathbf{n}\cdot\nabla\mathbf{u}\right)<0$, with $\mathbf{n}$ being the normal vector to the local iso-$\phi$ surface, correspond to the expansive and contractive behaviors, respectively.
	
	\subsection{Distance minimization problem}
	Seeking for a regularized level-set field, one ideally wants to enforce $\left|\nabla\phi\right|=1$.
	As first proposed by Li et al.~\cite{li_distance_2010}, this process can be based on the minimization of the potential function $P(\left| \nabla \phi \right|)$ with the property of
	\begin{equation}\label{eq:potential_argmin}
		\arg \, \min_s P(s) = 1.
	\end{equation}
	Having $P$ defined locally, the total virtual energy of the whole domain is obtained as
	\begin{equation}\label{eq:energy_functional}
		\mathcal{R}\left(\phi\right) = \int_\Omega P\left( \left| \nabla \phi \right| \right) d\Omega,
	\end{equation}
	and the sought for minimization problem becomes
	\begin{equation}\label{eq:minimization_condition}
		\frac{\partial \mathcal{R}\left(\phi\right)}{\partial \phi} = 0.
	\end{equation}
	
	\subsubsection{Variational formulation}
	The problem presented by Eq.~(\ref{eq:minimization_condition}) for energy functional $\mathcal{R}$, is equivalent to finding $\phi$ that satisfies 
	\begin{equation}\label{eq:minimization_condition2}
		d \mathcal{R}\left(\phi {;} \delta \phi \right) = 0 \quad \forall \: \delta \phi,
	\end{equation}
	where $\delta \phi$ denotes the variation in the level-set function.
	In order to calculate $d\mathcal{R}$, it is essential to take into account the directional variation of $\nabla \phi$ with respect to $\delta\phi$.
	To this end, the definition of the Gateaux differential can be used as 
	\begin{equation}\label{eq:Gateaux_derivative}
		d \mathcal{R}\left(\phi {;} \delta \phi \right) = \left. \frac{d}{d \varepsilon} \mathcal{R}\left(\phi + \varepsilon \delta \phi \right) \right|_{\varepsilon = 0}.
	\end{equation}
	Substituting the definition of the energy functional~(\ref{eq:energy_functional}) in Eq.~(\ref{eq:Gateaux_derivative}), one obtains
	\begin{equation}
		d \mathcal{R}\left(\phi {;} \delta \phi \right) = \int_\Omega P^\prime\left( \left| \nabla \phi \right| \right) \nabla \left( \delta \phi \right) \cdot \frac{\nabla \phi}{\left| \nabla \phi \right|} d\Omega,
	\end{equation}
	where 
	\begin{equation*}
		P^\prime\left( s \right) = \frac{d}{ds}P\left( s \right).
	\end{equation*}
	Consequently, the variational form of the minimization problem becomes
	\begin{equation}\label{eq:minimization_variational}
		\int_\Omega \frac{P^\prime\left( \left| \nabla \phi \right| \right)}{\left| \nabla \phi \right|} \nabla \psi \cdot \nabla \phi d\Omega = 0,
	\end{equation}
	with test function $\psi = \delta \phi$ representing the admissible variations.
	
	\subsubsection{Maintaining zero level-set}
	In order to retain the configuration of the surface defined as
	\begin{equation*}
		\Gamma(\phi_0) := \left\lbrace \mathbf{x} \in \Omega | \phi(\mathbf{x}) = \phi_0(\mathbf{x}) \right\rbrace,
	\end{equation*}
	the total energy functional can be augmented with an external energy function, $\mathcal{E}$, whose minimum occurs at $\Gamma(\phi_0)$~\cite{li_distance_2010}. 
	As pointed out by Basting and Kuzmin~\cite{basting_minimization-based_2013}, a simple but effective choice for $\mathcal{E}$ is
	\begin{equation}\label{eq:external_energy_functional}
		\mathcal{E}\left(\phi, \phi_0\right) = \frac{1}{2} \int_{\Gamma(\phi_0)} \left(\phi - \phi_0\right)^2 d\Gamma.
	\end{equation}
	This external energy can be introduced into the minimization problem, Eq.~(\ref{eq:minimization_condition}), as a penalty term
	\begin{equation}\label{eq:minimization_condition_augmented}
		\frac{\partial}{\partial \phi}\left[ \mathcal{R}\left(\phi\right) + \alpha \mathcal{E}\left(\phi, \phi_0\right) \right] = 0,
	\end{equation}
	where $\alpha$ is the penalty coefficient.
	The resulting variational form of the minimization problem reads
	\begin{equation}\label{eq:minimization_variational_augmented}
		\int_\Omega \frac{P^\prime\left( \left| \nabla \phi \right| \right)}{\left| \nabla \phi \right|} \nabla \psi \cdot \nabla \phi d\Omega + \alpha \int_{\Gamma(\phi_0)} \psi \left(\phi - \phi_0\right) d\Gamma = 0.
	\end{equation}
	Alternatively, one can acquire the Lagrange multiplier approach~\cite{adams_high-order_2019} to enforce the constraint on the zero level-set surface.
	
	It is worth mentioning that Eq.~(\ref{eq:minimization_variational_augmented}) characteristically presents a nonlinear elliptic problem with the diffusion coefficient of
	\begin{equation}\label{eq:diffusion_coefficient_definition}
		d\left( \left| \nabla \phi \right| \right) = \frac{P^\prime\left( \left| \nabla \phi \right| \right)}{\left| \nabla \phi \right|}.
	\end{equation}
	Nevertheless, solving Eq.~(\ref{eq:minimization_variational_augmented}) does not require the introduction of any boundary conditions and can readily be linearized and further discretized to obtain the linear system of equations for nodal values of $\phi$ as the unknowns.
	Following Basting and Kuzmin~\cite{basting_minimization-based_2013}, the linearization can be based on a fixed-point algorithm, for which the $m$-th iteration step reads
	\begin{dmath}\label{eq:fixed-point}
		\int_\Omega \nabla \psi \cdot \nabla \phi^{(m+1)} d\Omega + \alpha \int_{\Gamma(\phi_0)} \psi \left(\phi^{(m+1)} - \phi_0\right) d\Gamma = \int_\Omega \left[ 1 - d\left( \left| \nabla \phi^{(m)} \right| \right) \right]\nabla \psi \cdot \nabla \phi^{(m)} d\Omega.
	\end{dmath}
	Moreover, it should be noted that in most of the applications, the preservation of the configuration of the zero level-set surface is of main importance, i.e. $\phi_0=0$ in the above equations. Therefore, in the following for more simplicity, $\Gamma$ represents the zero level-set surface.
	
	\subsubsection{Objective potential}
	The satisfactory performance of the described minimization problem for the re-initialization of the level-set function is centrally determined by the proper definition of the potential function.
	As elaborated in Eq.~(\ref{eq:potential_argmin}), $P\left( \left| \nabla \phi \right| \right)$ has the essential property of exhibiting a (local) minimum at $\left| \nabla \phi \right|=0$.
	According to this property, one can characterize the diffusion coefficient~(\ref{eq:diffusion_coefficient_definition}) to tend to zero as $\left| \nabla \phi \right| \rightarrow 1$.
	Accordingly, one trivial option for $P$ is
	\begin{equation}\label{eq:potential_1}
		P\left( s \right) = \frac{1}{2}\left( 1 - s \right)^2,
	\end{equation}
	which gives
	\begin{equation}\label{eq:diffusion_coeff_1}
		d\left( s \right) = 1 - \frac{1}{s}.
	\end{equation}
	It is easily seen that while the problem is well defined for $\left| \nabla \phi \right| \geq 1$ choosing this potential function, $d\left( \left| \nabla \phi \right| \right) \rightarrow -\infty$ as $\left| \nabla \phi \right| \rightarrow 0$.
	Considering the computational difficulties associated with a large negative (anti-)diffusion coefficient, it is favorable to modify the definition of $P$ for $\left| \nabla \phi \right| < 1$.
	
	To cope with this issue, trigonometric~\cite{li_distance_2010} and polynomial~\cite{basting_minimization-based_2013} double-well functions are proposed that pose a second local minimum at $\left| \nabla \phi \right| = c$, where $0 < c < 1$.
	However, this treatment is not suitable for general level-set applications without any guarantee for $c < \left| \nabla \phi \right|$.
	As a more practical treatment, one can introduce a tailor-made polynomial that keeps the anti-diffusivity finite for the whole range of $\left| \nabla \phi \right| < 1$.
	Satisfying the essential conditions of $d(1)=0$, and keeping $d<0$ for $\left| \nabla \phi \right| <1$, a viable option is to define
	\begin{equation}\label{eq:adams_diffusion_coefficient}
		d\left( s \right) = 
		\begin{cases}
			1 - \frac{1}{s}, & \text{if} \: s \geq 1\\
			s - 1, & \text{otherwise}
		\end{cases}
	\end{equation}
	This coefficient leads to the definition of the potential function proposed by Adams et al.~\cite{adams_high-order_2019} as
	\begin{equation}\label{eq:adams_potential}
		P\left( s \right) = 
		\begin{cases}
			\frac{1}{2}\left( 1 - s \right)^2, & \text{if} \: s \geq 1\\
			\frac{1}{6} + \frac{s^3}{3} - \frac{s^2}{2}, & \text{otherwise}
		\end{cases}
	\end{equation}
	
	In this work, a further relaxed definition for the anti-diffusive coefficient in the range of $\left| \nabla \phi \right| <1$ is proposed reading
	\begin{equation}\label{eq:proposed_diffusion_coefficient}
		d\left( s \right) = 
		\begin{cases}
			1 - \frac{1}{s}, & \text{if} \: s \geq 1\\
			\frac{1}{2 - s} - 1, & \text{otherwise}
		\end{cases}
	\end{equation}
	
\begin{figure}[!htb]
		\centering
		\includegraphics[width=0.7\textwidth]{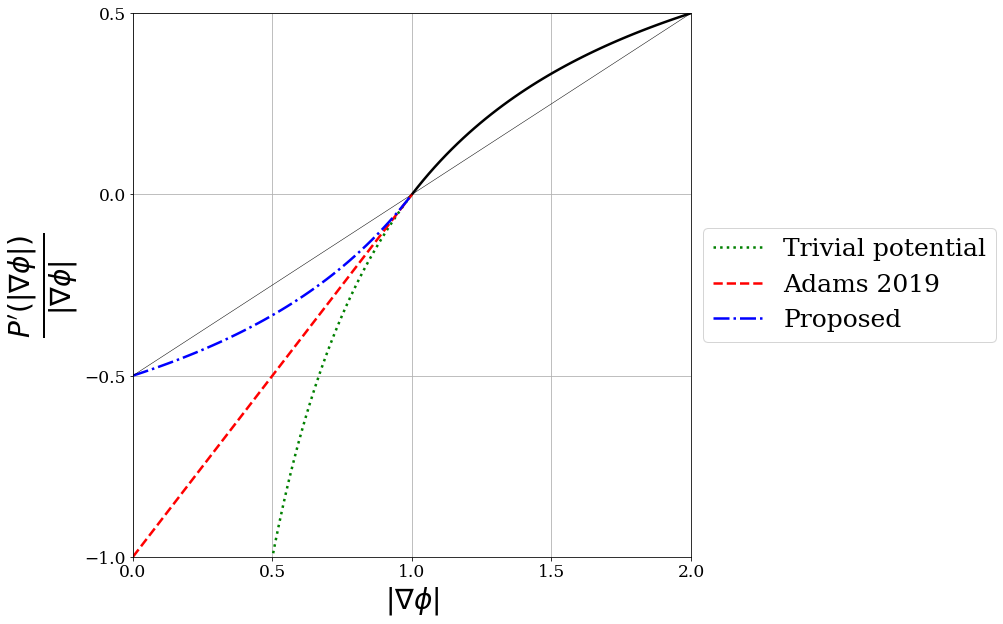}
		\caption{(Anti-)diffusion coefficient as a function of $\left| \nabla \phi \right|$ for different potentials.}
		\label{fig:diffusion_functionals}
	\end{figure}

	The rate of the variation of $d$ by changing $\left| \nabla \phi \right|$ is the same on both sides of the inflection point at $\left| \nabla \phi \right|=1$ (see Fig.~\ref{fig:diffusion_functionals}).
	This is obtained by reflecting $d(s)$ for $1 \leq s \leq 2$ around $(1, 0)$.
	Numerical experiments showed that this would slightly improve the rate of convergence of the problem presented in Eq.~(\ref{eq:fixed-point}). 
	
	\subsection{Elliptic reconstruction scheme}
	A major drawback of the optimization-based level-set re-initialization approach is its low rate of convergence.
	This issue is exacerbated in cases with a significant local reduction of $\left| \nabla \phi \right|$ with respect to the optimal value of unity.
	Therefore, the performance of the method can be dramatically enhanced if $\phi$ was initially recalculated in a way that $\left| \nabla \phi \right|$ remains reasonably close to unity, e.g., $0.5 < \left| \nabla \phi \right| < 2$, within the computational domain.
	
	In this work, the proposed scheme for the reconstruction of $\phi$ is based on the solution of 
	\begin{equation}\label{eq:reconstruction_equation}
		\nabla \cdot \nabla \phi = q\left(\mathbf{x}\right),
	\end{equation}
	with a prescribed source term $q$.
	This equation is subject to Neumann conditions defined on the outer boundaries of the domain, $\partial \Omega$. 
	The corresponding weak form of Eq.~(\ref{eq:reconstruction_equation}) that includes the penalization term for preserving zero level-set reads
	\begin{dmath}\label{eq:reconstruction_equation_variational}
		\int_\Omega \nabla \psi \cdot \nabla \phi d\Omega - \int_{\partial \Omega} \mathbf{n} \cdot \nabla \phi d \partial \Omega + \alpha \int_{\Gamma} \psi \phi d\Gamma + \int_\Omega \psi q\left(\mathbf{x}\right) d\Omega = 0.
	\end{dmath}
	Considering Eq.~\ref{eq:reconstruction_equation} and taking into account that the ideal solution would feature with $\left| \nabla \phi \right| = 1$, one can obtain the admissible definition of $q\left(\mathbf{x}\right) = \nabla \cdot \left( \nabla \phi/\left| \nabla \phi \right| \right)$.
	In other words, source term $q\left(\mathbf{x}\right)$ in Eq.~\ref{eq:reconstruction_equation} is equal to the local curvature of the iso-$\phi$ surface passing through $\mathbf{x}$.
	
	\subsubsection{Boundary condition}\label{sec:blind_spot}
	As presented in Eq.~(\ref{eq:reconstruction_equation_variational}), the successful application of the proposed level-set reconstruction scheme is subject to the natural imposition of the (estimated) gradient of the level-set function on the boundaries of the computational domain.
	Nonetheless, $\mathbf{n}\cdot\nabla\phi$ should be consistently evaluated not to disturb the configuration of the contact line.
	In this work, this boundary term is calculated using the geometrical definition of the regularized level-set function.
	The technique discussed below can also be readily utilized to improve any geometrical level-set re-initialization technique (e.g., see~\cite{henri_geometrical_2022}).
	
\begin{figure}[!htb]
		\centering
		\includegraphics[width=0.4\textwidth]{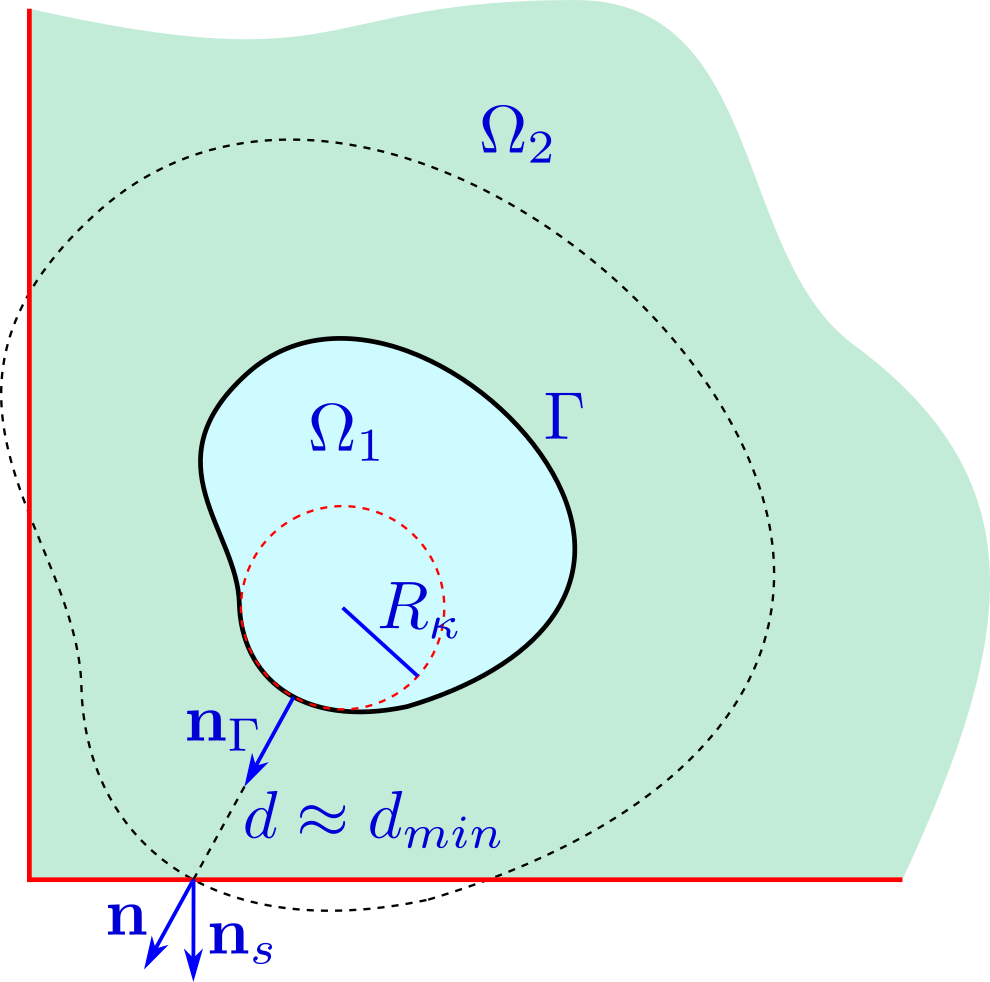}
		\caption{Reproduction of the boundary condition for a free droplet.}
		\label{fig:boundary_free_droplet}
	\end{figure}

\begin{figure}[!htb]
		\centering
		\includegraphics[width=0.4\textwidth]{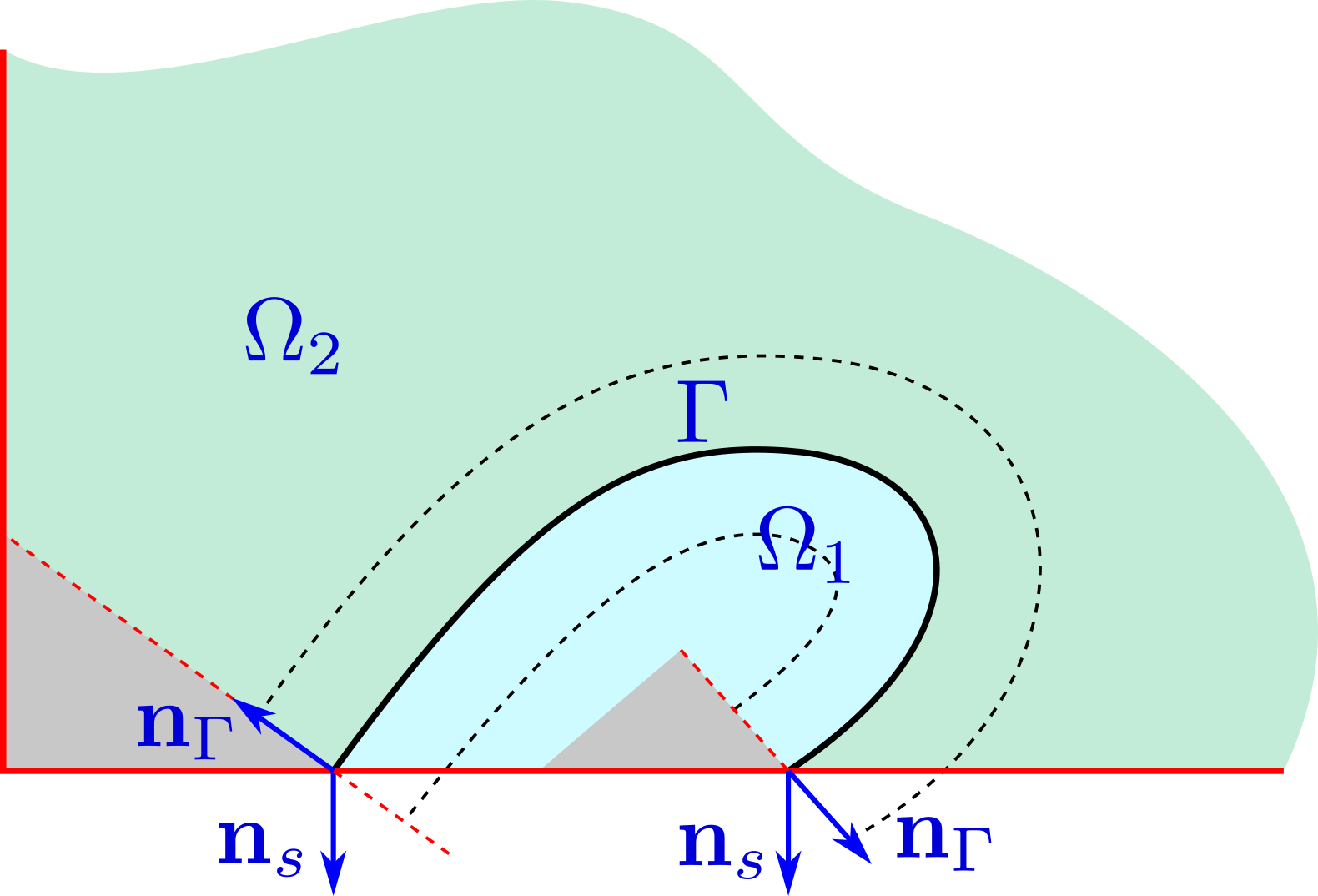}
		\caption{Blind spots in the level-set domain adjacent to a droplet cut by a substrate.}
		\label{fig:boundary_cut_droplet}
	\end{figure}

	Knowing that for an idea level-set function, the normal vector to the phase interface reads $\mathbf{n}_\Gamma = \nabla\phi$, one can project $\mathbf{n}_\Gamma$ onto the outer boundaries to estimate the corresponding boundary condition.
	As illustrated in Fig.~\ref{fig:boundary_free_droplet}, at each boundary node, the projected normal vector originates from the nearest point on the phase interface (the preserved zero level-set).
	Even though this simple procedure works fine in the absence of contact line, once the zero level-set contour is cut by an outer boundary, a specific treatment is necessary for estimating the boundary condition in the blind spots shown in Fig.~\ref{fig:boundary_cut_droplet}.
	Such a condition frequently occurs in different applications, for example, in the case of a sessile droplet on a solid substrate.
	
\begin{figure}[!htb]
		\centering
		\includegraphics[width=0.4\textwidth]{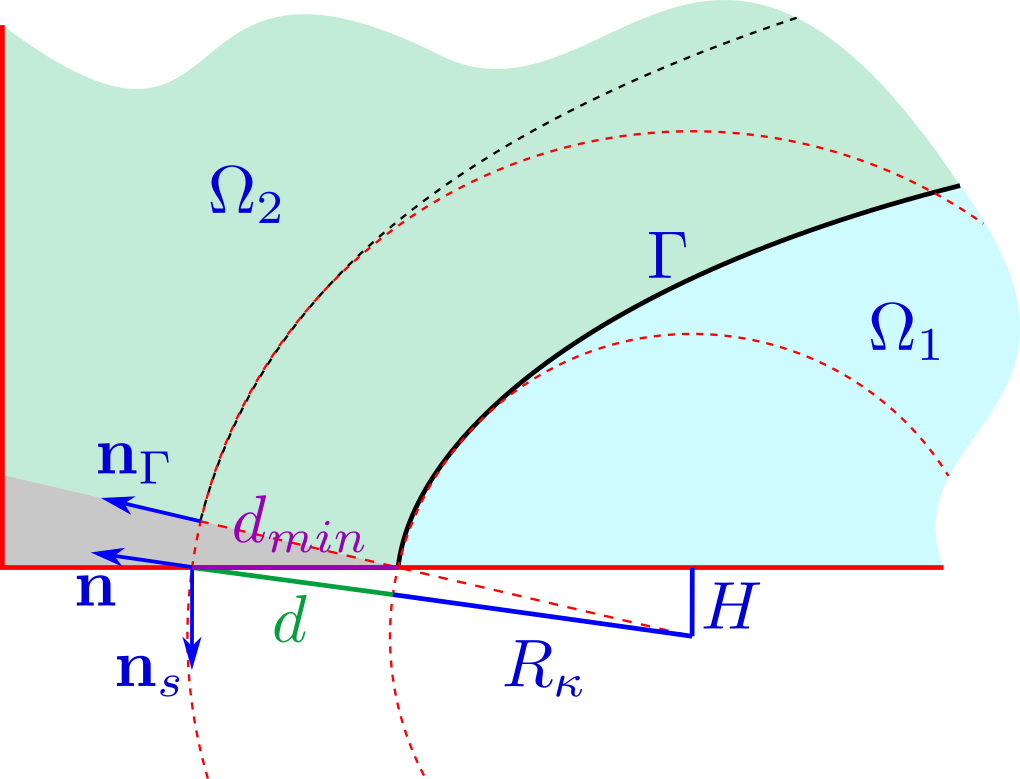}
		\caption{Reproduction of the boundary condition in the blind spot adjacent to the cut interface.}
		\label{fig:boundary_cut_droplet_blind_spot_adjacent}
	\end{figure}

\begin{figure}[!htb]
		\centering
		\includegraphics[width=0.4\textwidth]{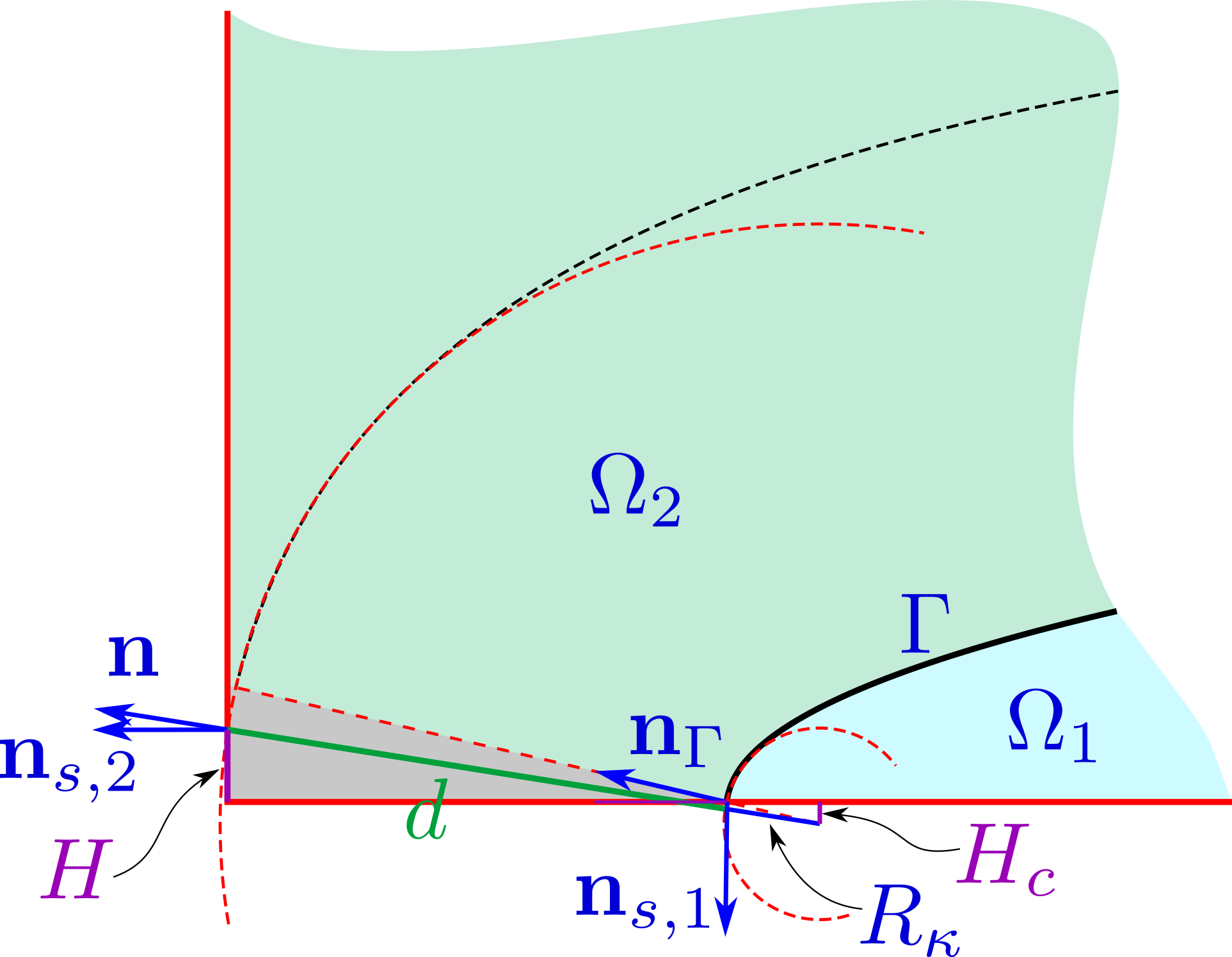}
		\caption{Reproduction of the boundary condition in the blind spot on the side boundaries.}
		\label{fig:boundary_cut_droplet_blind_spot_side}
	\end{figure}

	In order to treat the blind spot, the associated boundary nodes are divided into two different groups; those lying on the cut plane (see Fig.~\ref{fig:boundary_cut_droplet_blind_spot_adjacent}) and those lying on the boundaries perpendicular to the cut plane (see Fig.~\ref{fig:boundary_cut_droplet_blind_spot_side}).
	The treatment here is based on hypothetically extending the zero level-set contour assuming a constant curvature behind the cut plane.
	The procedure is illustrated in Figs.~\ref{fig:boundary_cut_droplet_blind_spot_adjacent} and~\ref{fig:boundary_cut_droplet_blind_spot_side} for both groups of the boundary nodes in the blind spot.
	In this way, one can approximate the normal gradient of the level-set function within the blind spot as
	\begin{equation}
		\mathbf{n}_s \cdot \mathbf{n} \approx - \sin\left({\frac{- R_\kappa \left( \mathbf{n}_s \cdot \mathbf{n}_\Gamma \right)}{R_\kappa + d}}\right),
	\end{equation}
	and
	\begin{equation}
		\mathbf{n}_{s,2} \cdot \mathbf{n} \approx - \sqrt{ \left( R_\kappa + d \right)^2 + \left[ H - R_\kappa (\mathbf{n}_{s,1} \cdot \mathbf{n}_\Gamma) \right]^2 },
	\end{equation}
	for the first and second groups of the boundary nodes, respectively.
	
	It is worth emphasizing that while solving Eq.~(\ref{eq:reconstruction_equation_variational}) with the local curvature as the source term results in a relatively regularized $\phi$, the absence of the aforementioned minimization problem (Eq.~(\ref{eq:minimization_condition_augmented})) offers no assurance of achieving $\left| \nabla \phi \right| \approx 1$.
	The approximations made for estimating the boundary condition term can also potentially increase the deviation from the ideal level-set function.
	On the other hand, the minimization procedure governed by Eq.~(\ref{eq:minimization_condition_augmented}) does not require the introduction of any boundary conditions, and therefore, can be seen as a robust corrector for the proposed level-set re-construction scheme with a low sensitivity to the possible errors in the estimation of the boundary condition term.
	
	\subsection{Element splitting and preserving contact line}\label{elem_split}
	As elaborated earlier, the imposition of the zero level-set constraint is one of the most important requirements for the success of the level-set re-initialization process.
	In this regard, for the described class of elliptic approaches, one needs to accurately calculate the corresponding penalty terms.
	To this end, not having ready access to the zero level-set surface on the fixed computational mesh, one can acquire a mechanism to project the terms into the nodal contributions~\cite{xue_new_2021}.
	However, the more versatile and straightforward approach would be based on the element-splitting process described in this section.
	
\begin{figure}[!htb]
		\centering
		\includegraphics[width=0.9\textwidth]{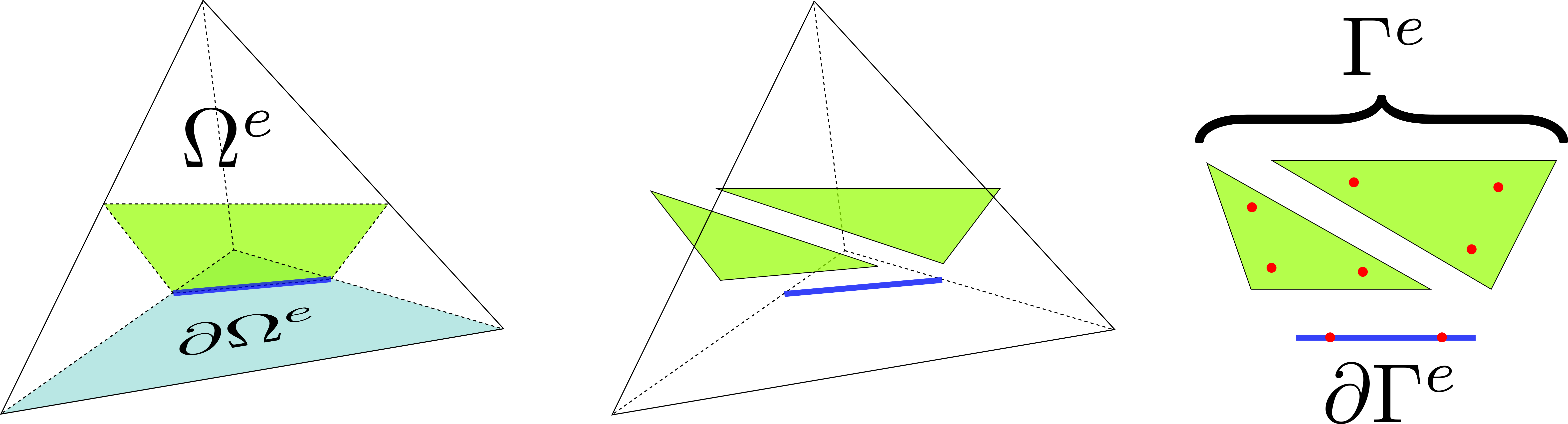}
		\caption{Generation of the sub-element entities for the calculation of the interface and contact line preserving penalty terms. In this figure, the portion of the outer boundary of the domain that coincides with element $\Omega^e$ is shaded and denoted by $\partial \Omega^e = \partial\Omega \cap \Omega^e$. The points associated with the Gauss quadrature are also shown as red solid dots.}
		\label{fig:tetrahedron_interface}
	\end{figure}

	The splitting of the cut elements starts with the introduction of the corners of the phase interface ($\phi=0$) plane on the edges of the element.
	For the linear tetrahedral elements used in this work, depending on the configuration, this leads to either a quadrilateral or a triangular cut plane representing the elemental portion of the phase interface.
	In the case of the quadrilateral shape, as illustrated in Fig.~\ref{fig:tetrahedron_interface}, the cut plane is further split into two triangles.
	Gauss quadrature is then used to integrate the contribution of the penalty term ensuring the preservation of the location of the phase interface. 
	
	As previously mentioned, in droplet dynamics simulations, accurate capturing of the contact angle plays an essential role.
	This requirement urges the preservation of the location of the contact line while the zero level-set contour is fixed.
	With this aim, one can simply extend the penalty terms appearing in Eqs.~(\ref{eq:fixed-point}) and~(\ref{eq:reconstruction_equation_variational}) to read
	\begin{dmath}\label{eq:fixed-point-contact line}
		\int_\Omega \nabla \psi \cdot \nabla \phi^{(m+1)} d\Omega + \alpha_\Gamma \int_{\Gamma} \psi \phi^{(m+1)} d\Gamma + \alpha_{\partial\Gamma} \int_{\partial\Gamma} \psi \phi^{(m+1)} d\partial \Gamma = \int_\Omega \left[ 1 - d\left( \left| \nabla \phi^{(m)} \right| \right) \right]\nabla \psi \cdot \nabla \phi^{(m)} d\Omega,
	\end{dmath}
	and
	\begin{dmath}\label{eq:reconstruction_equation_variational-contact line}
		\int_\Omega \nabla \psi \cdot \nabla \phi d\Omega - \int_{\partial \Omega} \mathbf{n} \cdot \nabla \phi d \partial \Omega + \alpha_\Gamma \int_{\Gamma} \psi \phi d\Gamma + \alpha_{\partial\Gamma} \int_{\partial\Gamma} \psi \phi d\partial \Gamma + \int_\Omega \psi q\left(\mathbf{x}\right) d\Omega = 0,
	\end{dmath}
	where $\partial \Gamma$ denotes the contact line, which is represented by the cut in the zero level-set contour.
	Coefficients $\alpha_\Gamma$ and $\alpha_{\partial\Gamma}$ are chosen large enough to assure the fixation of the phase interface (including the contact line).
	In this work, we have presented our formulation according to the conventional penalty method to maintain generality and enhance readability. However, for the sake of improved numerical stability and efficiency, our numerical implementation is based on Nitsche's scheme~\cite{nitsche_uber_1971}.
	
	In summary, our proposed re-initialization method is a two-step process: in the first step, the level-set function is fully reconstructed using Eq.~\ref{eq:reconstruction_equation_variational-contact line}, ensuring a smooth initial state. Subsequently, re-initialization is achieved through an iterative optimization procedure based on an elliptic equation Eq.~\ref{eq:fixed-point-contact line}, by incorporating a feasible objective potential (Eq.~\ref{eq:proposed_diffusion_coefficient}), and considering element splitting approach of Section~\ref{elem_split} within. Notably, special treatment, detailed in Section~\ref{sec:blind_spot}, is applied when handling blind spots in boundary condition calculations. This two-step approach ensures accurate preservation of the zero level-set configuration and mitigates the issues commonly encountered in traditional re-initialization methods.
	
	\section{Results}
	
	The method has been implemented within \textit{Kratos Multiphysics} framework, an in-house Open Source C++ object-oriented Finite Element platform~\cite{ferrandiz_kratosmultiphysicskratos_2023, dadvand_object-oriented_2010}. To simulate the fluid dynamics, we employ the enriched Finite Element model introduced by the authors in previous works~\cite{hashemi_enriched_2020,hashemi_three_2021}. Furthermore, the dynamics of the contact line is modelled using a combination of the linear molecular kinetic theory and hydrodynamic theory, as previously detailed in our earlier publication~\cite{hashemi_toward_2021}.
	
	Unless specified otherwise, the properties of the liquid and gas phases correspond to those of water and air, respectively. These properties are set as follows: water density, $\rho_l = 1000~\text{kg}/\text{m}^3$; water dynamic viscosity, $\mu_l = 0.001$ Pa s; air density, $\rho_g = 1~\text{kg}/\text{m}^3$; air dynamic viscosity, $\mu_g = 0.00001$ Pa s; and the surface tension of the water-air interface, $\gamma = 0.072$ N/m. Gravity is consistently defined as $g = 9.81~\text{m}/\text{s}^2$ in all test cases, except for the Oscillating Droplet scenario (Sec.~\ref{sec:OsciDrop}), where it is neglected. The simulations are carried out within a cubic domain measuring 1 mm $\times$ 1 mm $\times$ 1 mm, with the exception of the Droplet in Channel (Sec.~\ref{sec:DropletChannel}), where the domain dimensions are different.
	
	
	Three key capabilities and goals are aimed to be achieved by our approach: 1) maintaining the magnitude of the level-set gradient at unity, i.e., $|\nabla \phi_i| = 1$, 2) accurately preserving the liquid-gas interface, and 3) ensuring the preservation of the contact line. To test the method's performance in achieving these goals, appropriate error measures are defined for each objective.
	
	For the first goal, ensuring $|\nabla \phi_i| = 1$, the $L^1$-norm of the error in the magnitude of the level-set gradient, denoted as
	\begin{equation}
		L^1\text{-norm}(|\nabla \phi_i|) = \frac{1}{N_{\Omega^\prime}}\sum_{i=1}^{N_{\Omega^\prime}}(|\nabla \phi_i| - 1),
		\label{eq:l1norm}
	\end{equation}
	is utilized as a quantitative measure, where $N_{\Omega^\prime}$ is the total number of nodes in the subdomain $\Omega^\prime\subset\Omega$. This error metric allows the assessment of the method's capability in maintaining the correct magnitude of the gradient throughout the optimization process. It's important to note that our evaluation is not limited to the entire domain or \textit{full-band}, i.e., $\Omega^\prime\equiv\Omega$, but extends to a \textit{narrow-band} defined as $\Omega^\prime := \left\lbrace \mathbf{x} \in \Omega | -3h<\phi(\mathbf{x}) <+3h\right\rbrace$. This extension allows us to evaluate the method's performance in the close vicinity of the liquid-gas interface, where elements are intersected by the interface, necessitating element splitting.
	
	To evaluate the second desired capability, namely, accurate preservation of the interface $\Gamma$, the Chamfer Distance (CD) between the reinitialized interface, $\phi_0^{\text{reinit}}$, and the exact interface, $\phi_0^{\text{exact}}$, is employed. The Chamfer Distance is calculated as
	\begin{equation}
		\text{CD}(\Gamma) = \frac{1}{N_{\Gamma}} \sum_{i=1}^{N_{\Gamma}} \min_{\mathbf{x}_j \in \Gamma_{\text{exact}}} \|\mathbf{x}_i - \mathbf{x}_j\|,
		\label{eq:CD}
	\end{equation}
	where $N_{\Gamma}$ is the total number of nodes on the interface $\Gamma$, and $\mathbf{x}_i$ represents the position of the $i$-th node on $\Gamma$. Here, $\Gamma_{\text{exact}}$ represents the set of nodes on the exact interface.
	
	Similarly, the Hausdorff Distance (HD) is used as a metric to assess the method's ability to preserve the contact line $\partial \Gamma$, and can be computed as follows:
	\begin{equation}
		\text{HD}(\partial \Gamma) = \max \left( \sup_{\mathbf{x}_i \in \partial \Gamma_{\text{reinit}}} \min_{\mathbf{x}_j \in \partial \Gamma_{\text{exact}}} \|\mathbf{x}_i - \mathbf{x}_j\|, \sup_{\mathbf{x}_j \in \partial \Gamma_{\text{exact}}} \min_{\mathbf{x}_i \in \partial \Gamma_{\text{reinit}}} \|\mathbf{x}_i - \mathbf{x}_j\| \right),
		\label{eq:HD}
	\end{equation}
	where $\partial \Gamma_{\text{reinit}}$ and $\partial \Gamma_{\text{exact}}$ represent the sets of nodes on the reinitialized and exact contact line $\partial \Gamma$ respectively.
	
	\subsection{Verification}
	In this subsection, we verify the performance of the proposed level-set re-initialization method using two benchmark test cases: a) Spherical Droplet and b) Ellipsoid Droplet. Despite their apparent simplicity, these cases serve as fundamental building blocks to evaluate the method's generalizability to more complex scenarios. Additionally, the presence of analytical solutions enables direct comparison with numerical results, validating the accuracy and reliability of our proposed approach.
	
	\subsubsection{Spherical droplet}
	The first benchmark test case involves simulating a sessile spherical droplet within a fixed computational domain. This case is highly relevant as it allows us to explore the method's ability to handle sessile droplets and preserve the contact line, representing the three-phase (liquid-gas-solid) situation. The presence of the contact line is crucial, as it highlights the method's efficacy in dealing with complex interfacial dynamics during re-initialization while accurately preserving the zero level-set configuration.
	
	\begin{figure}[!htb]
		\begin{subfigure}{\textwidth}
			\centering 
			\subcaptionOverlay{\includegraphics[width=0.3\textwidth]{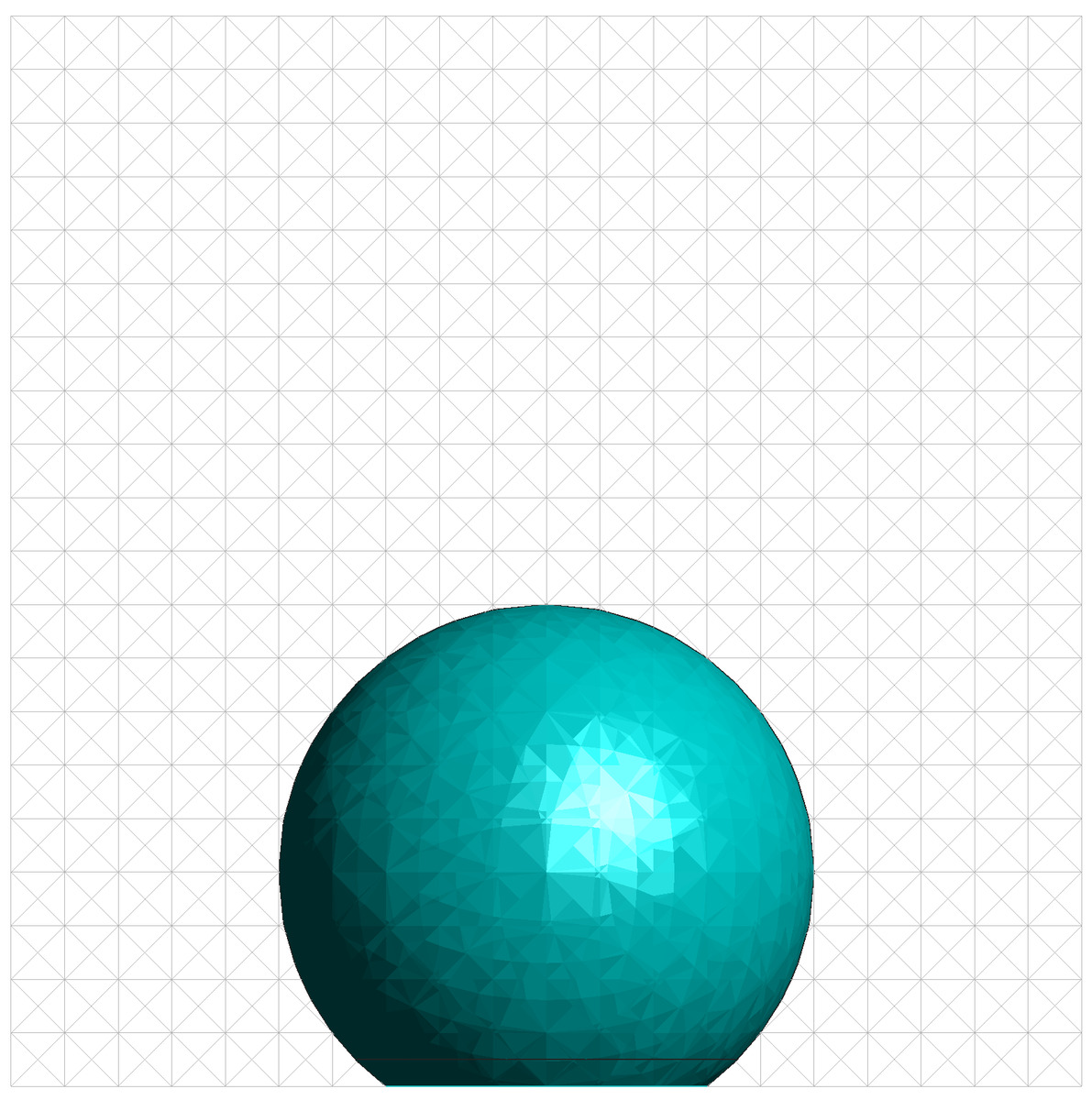}
				\includegraphics[width=0.3\textwidth]{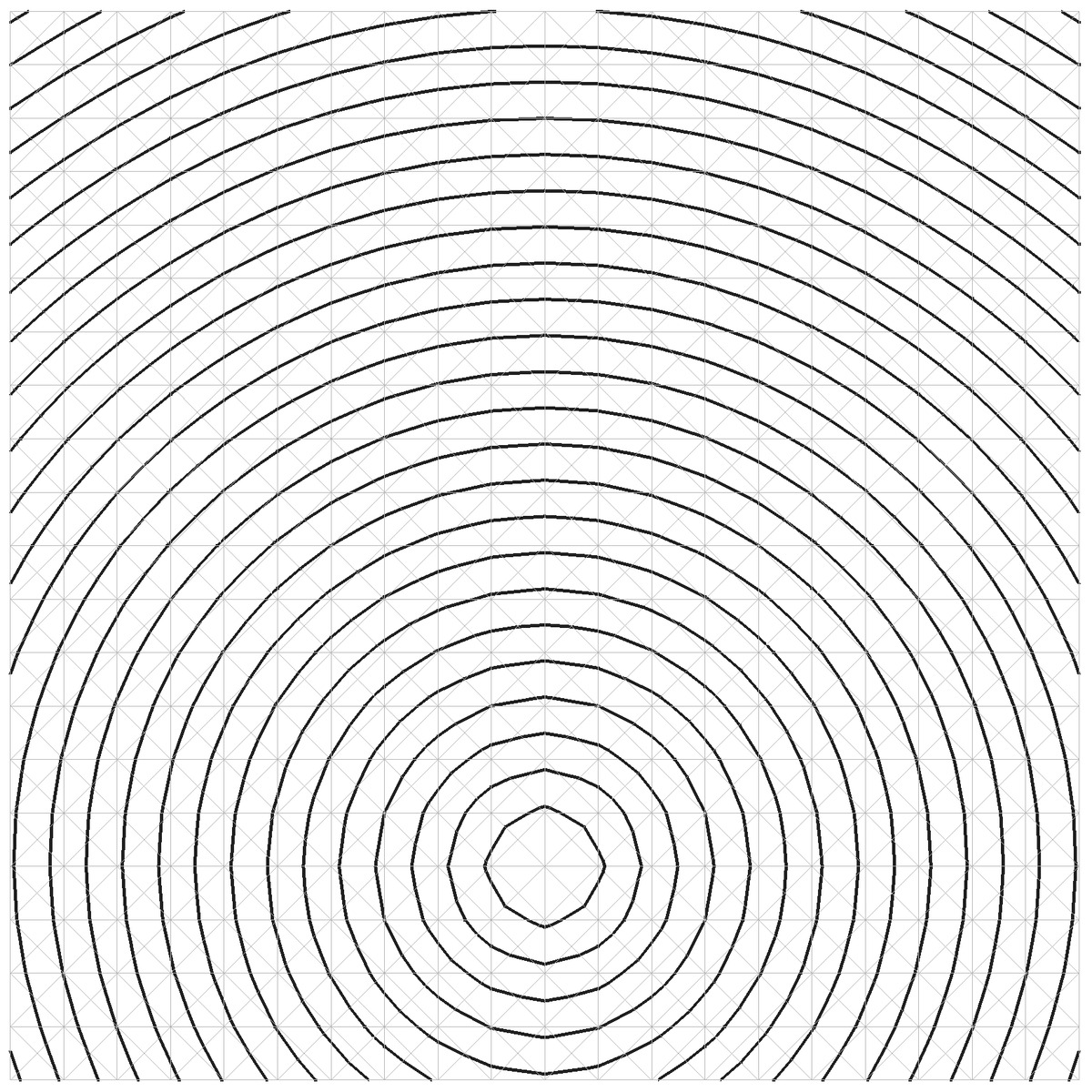}}
		\end{subfigure}    
		\hfill
		\begin{subfigure}{\textwidth}
			\centering
			\subcaptionOverlay{\includegraphics[width=0.3\textwidth]{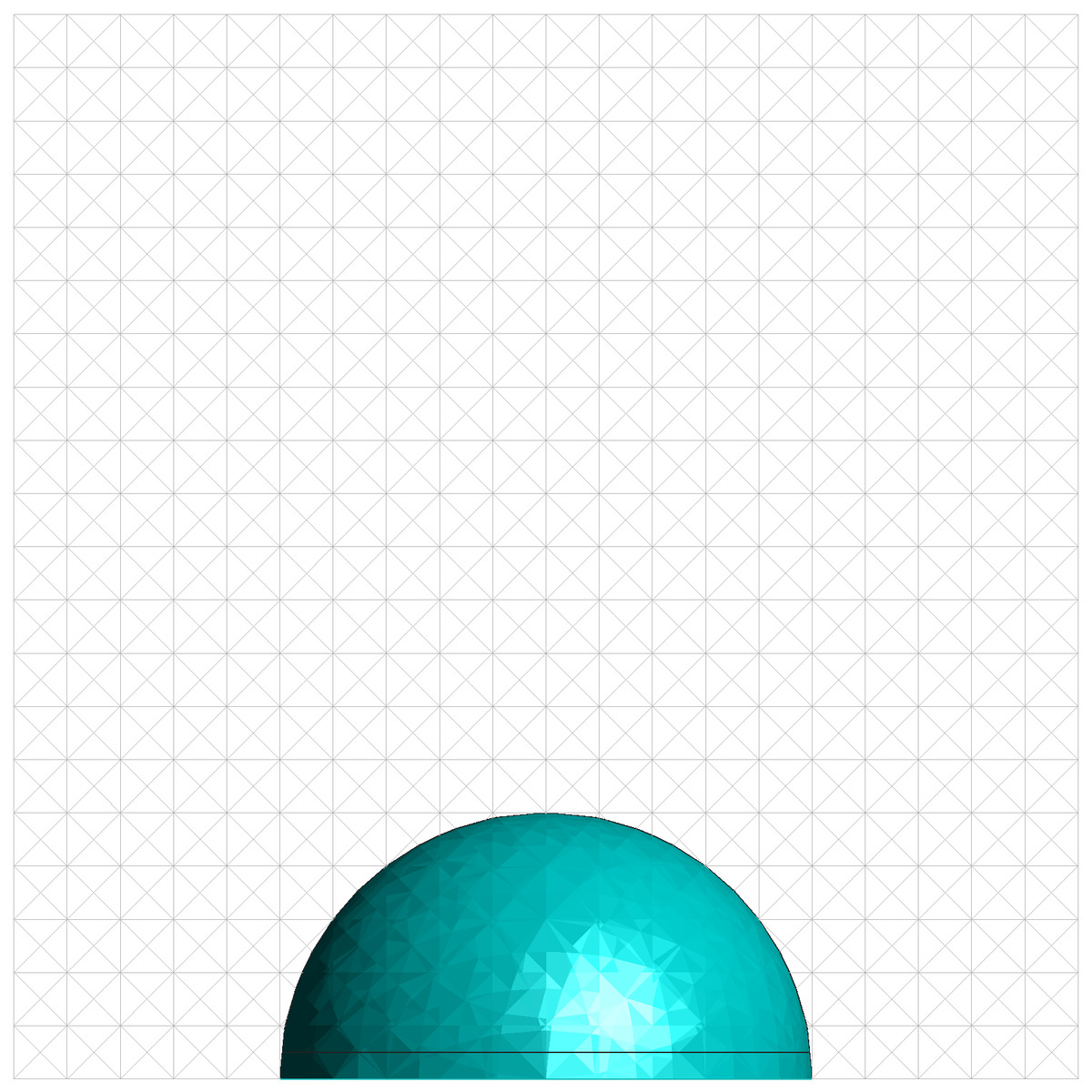}
				\includegraphics[width=0.3\textwidth]{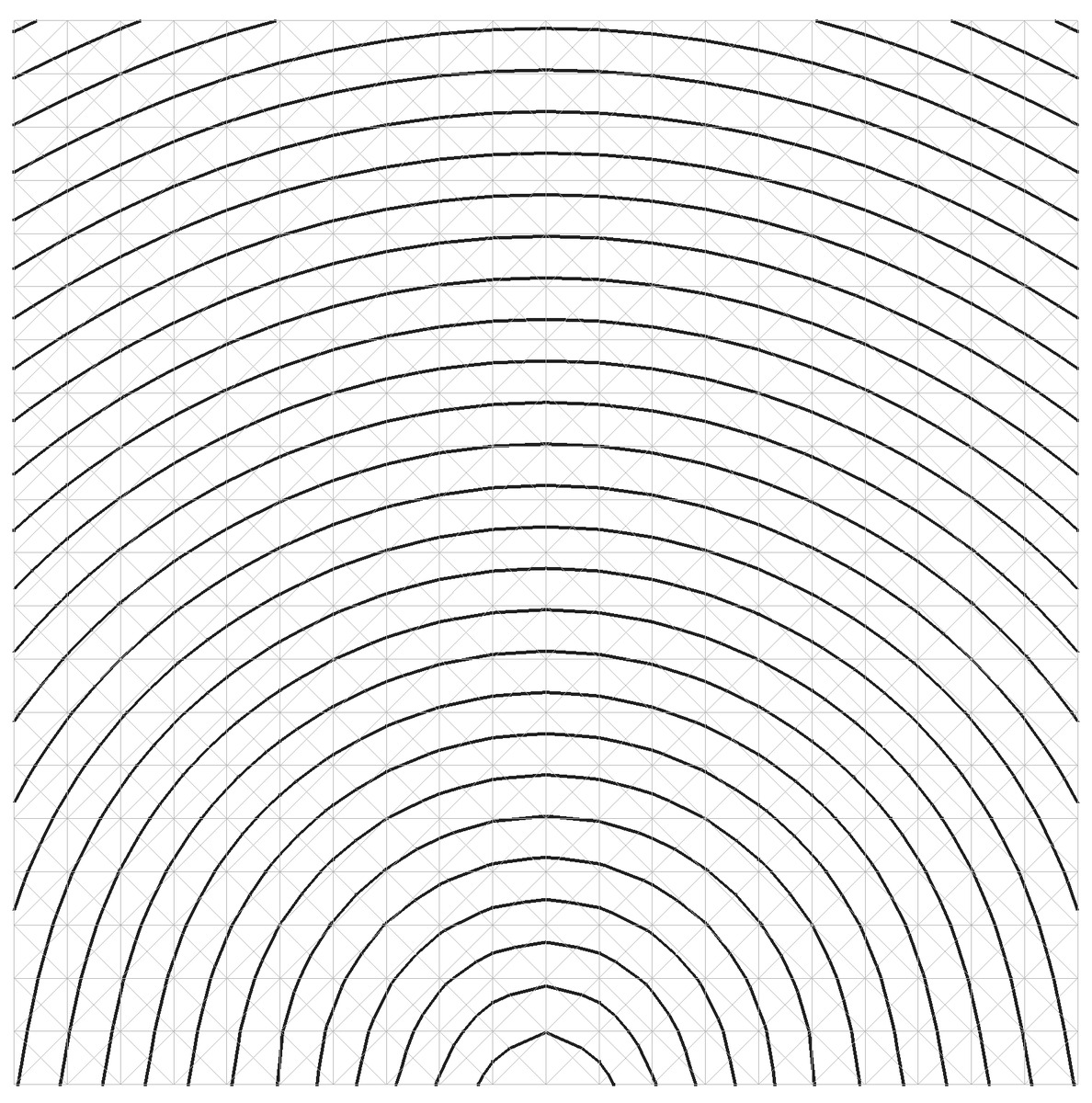}}
		\end{subfigure}
		\hfill
		\begin{subfigure}{\textwidth}
			\centering
			\subcaptionOverlay{\includegraphics[width=0.3\textwidth]{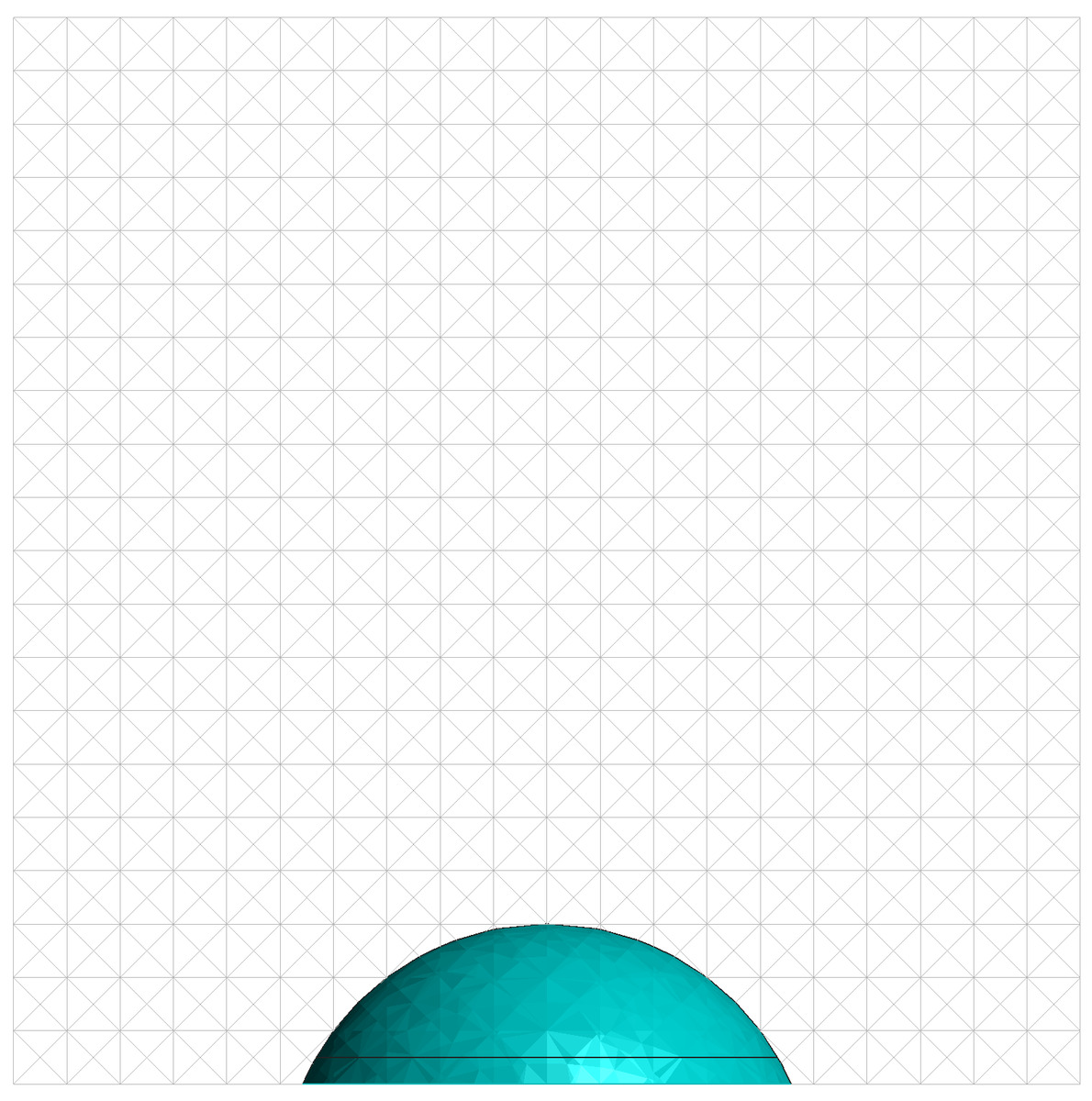}
				\includegraphics[width=0.3\textwidth]{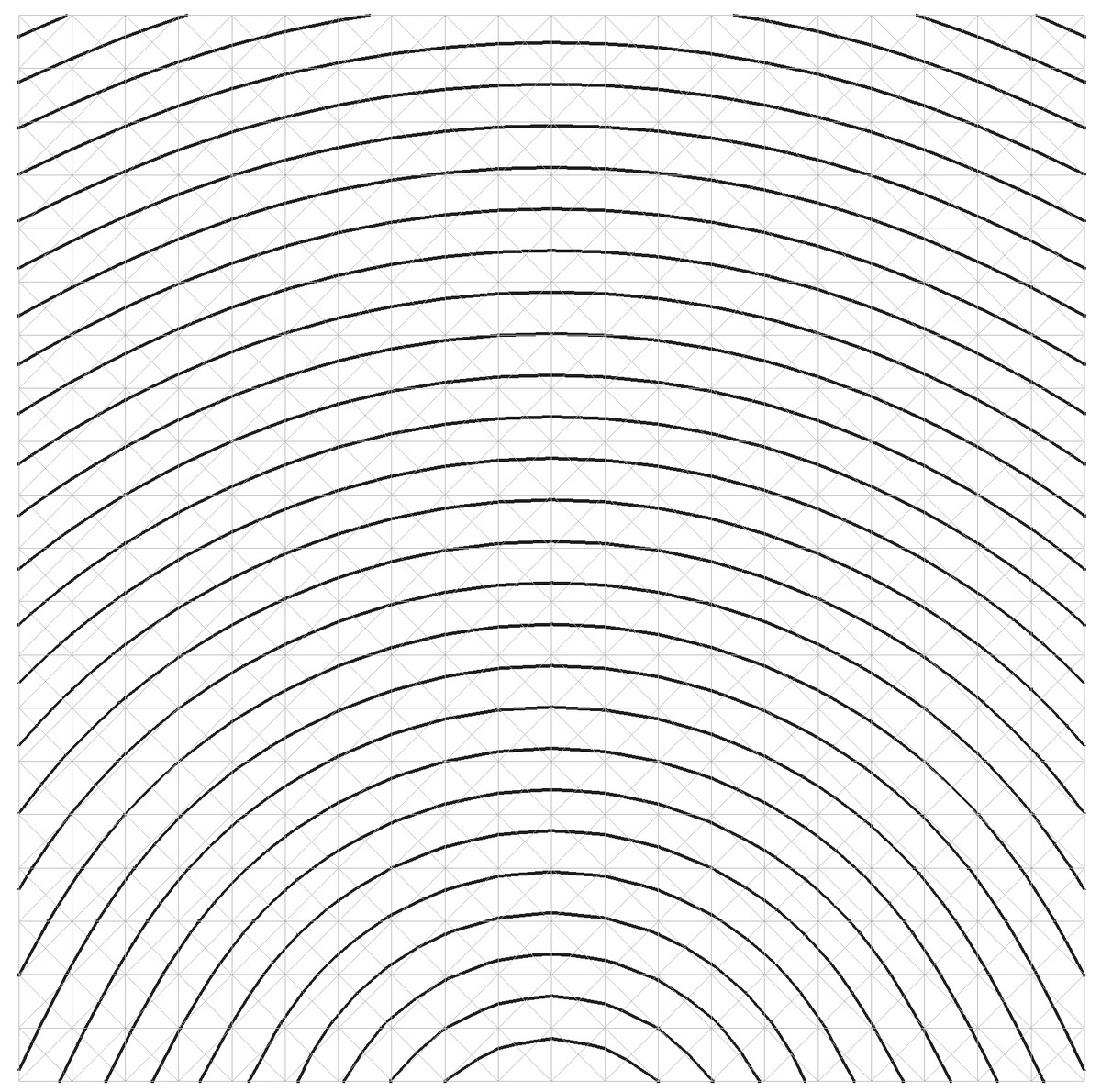}}
		\end{subfigure}
		\caption{Side views of three spherical droplets on a substrate with different contact angles (Obtuse angle, Right angle, and Acute angle) and same radius $R=250\mu$m (right column), as well as, the corresponding contours of iso-level-set (left column)}
		\label{fig:spherical}
	\end{figure}
	Side views of three spherical droplets on a substrate with different contact angles (Obtuse angle, Right angle, and Acute angle) and fixed radius $R=250\mu$m are presented in Fig.~\ref{fig:spherical}. The contours of iso-level-set are displayed for each case, providing a clear visualization of their regularity.
	
	\begin{figure}[!htb]
		\centering
		\includegraphics[width=0.8\textwidth]{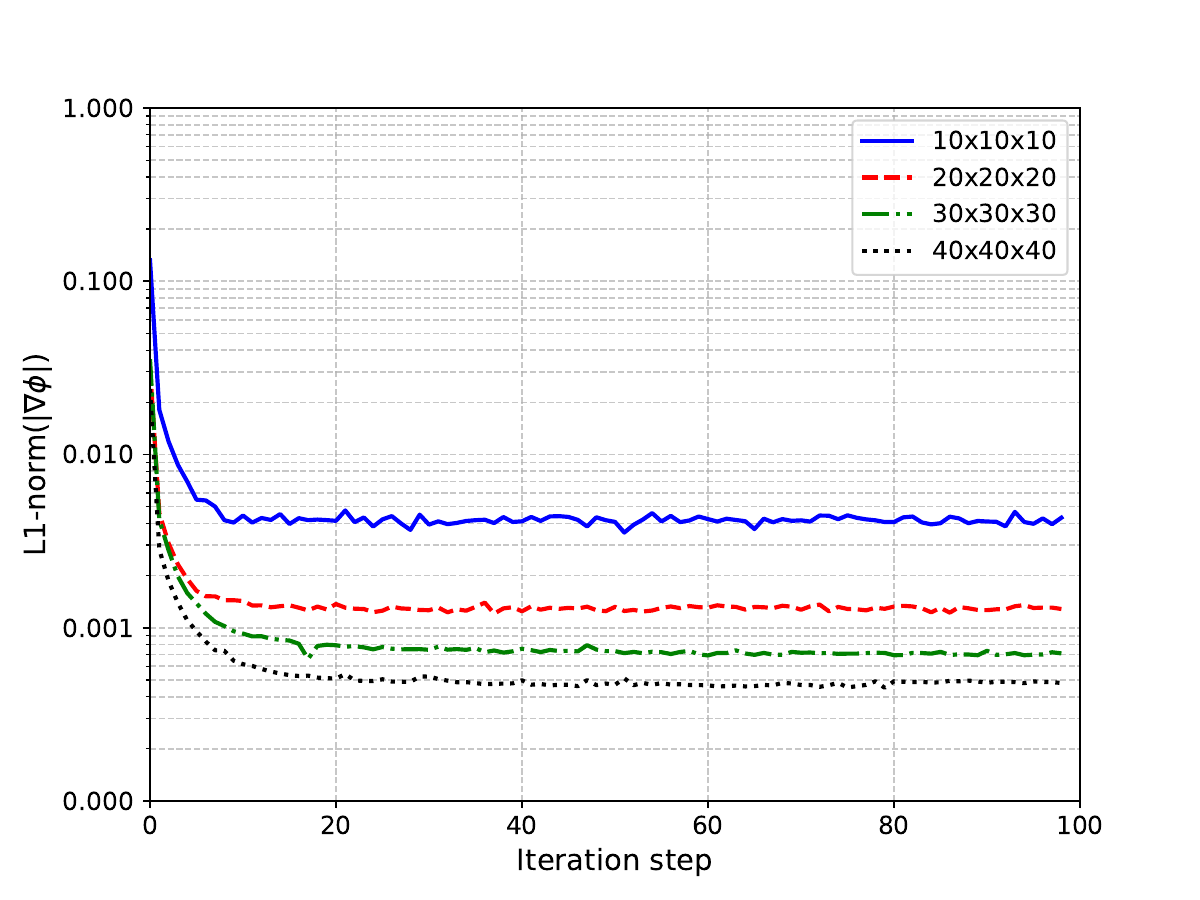}
		\caption{Evolution of the (average) $L^1$-norm of the error in the gradient of the level-set function by during the iterative optimization procedure.}
		\label{fig:sphere_iterations}
	\end{figure}
	The evolution of the (average) $L^1$-norm of the error in the gradient of the level-set function during the iterative optimization procedure is illustrated in Fig.~\ref{fig:sphere_iterations}. It reveals that the error rapidly decreases to its minimum within the initial iteration steps of the iterative optimization procedure. Notably, the plot displays consistent results for different mesh sizes, ranging from coarser to finer meshes. This robust convergence behavior demonstrates the effectiveness of the proposed method in achieving accurate and efficient optimization of the level-set function, regardless of the mesh resolution.
	
	\begin{figure}[!htb]
		\begin{subfigure}{\textwidth}
			\centering
			\makebox[0.5\textwidth][c]{\scriptsize\textbf{full-band}}%
			\makebox[0.5\textwidth][c]{\scriptsize\textbf{narrow-band}}%
			\subcaptionOverlay{\includegraphics[width=0.5\textwidth]{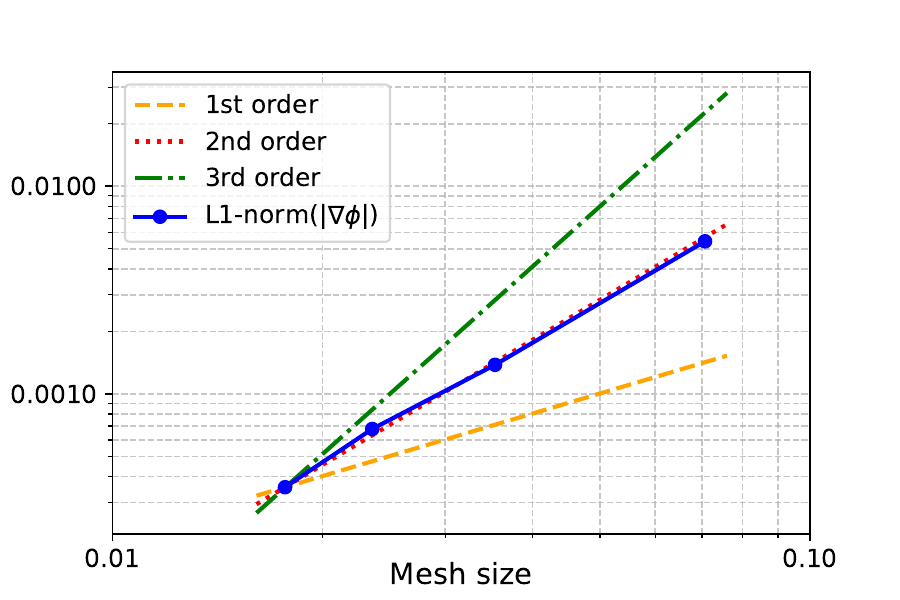}
				\includegraphics[width=0.5\textwidth]{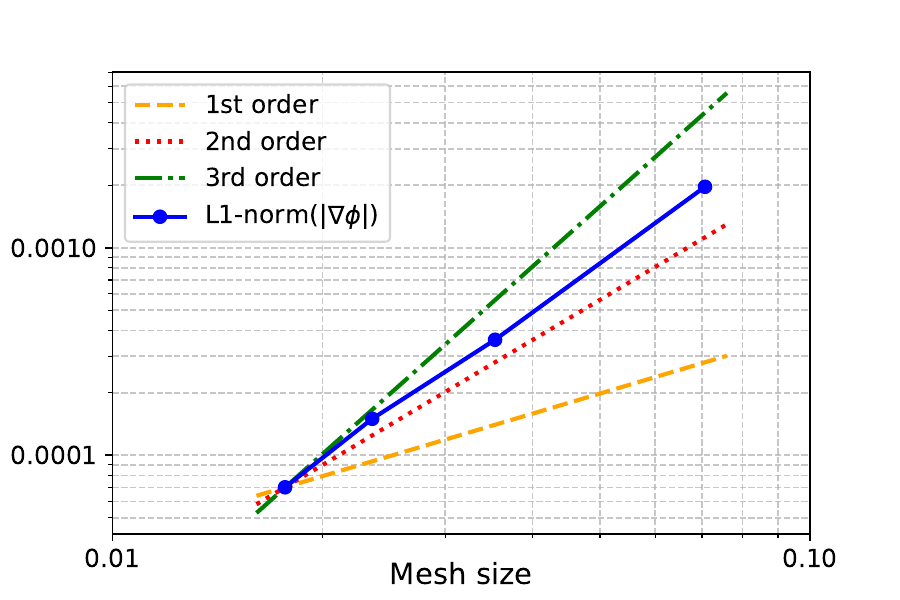}}
		\end{subfigure}    
		\hfill
		\begin{subfigure}{\textwidth}
			\centering
			\subcaptionOverlay{\includegraphics[width=0.5\textwidth]{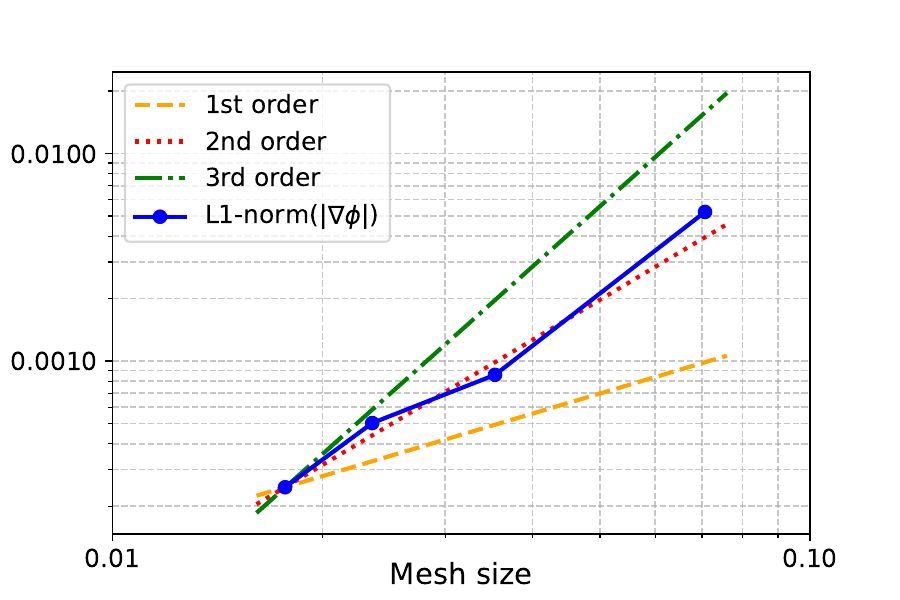}
				\includegraphics[width=0.5\textwidth]{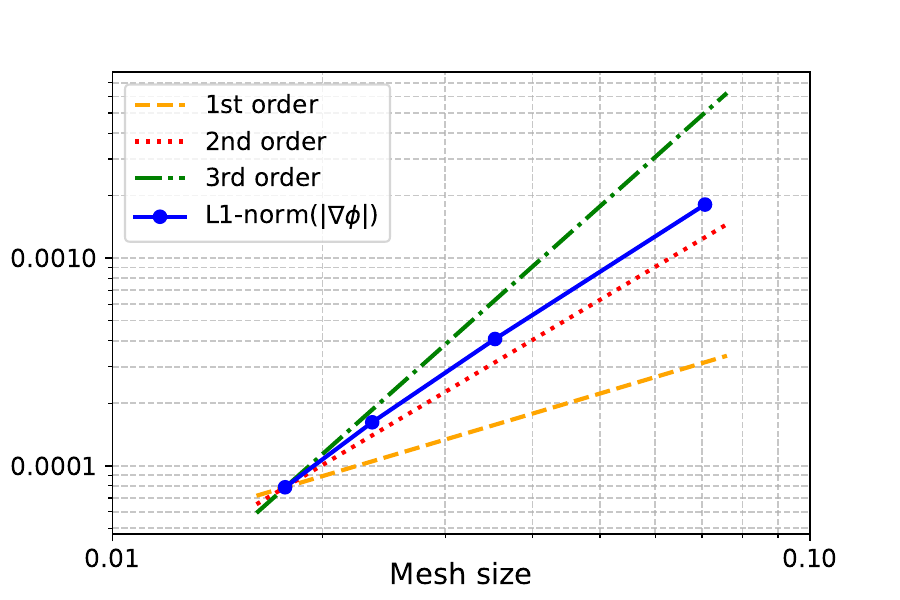}}
		\end{subfigure}
		\hfill
		\begin{subfigure}{\textwidth}
			\centering
			\subcaptionOverlay{\includegraphics[width=0.5\textwidth]{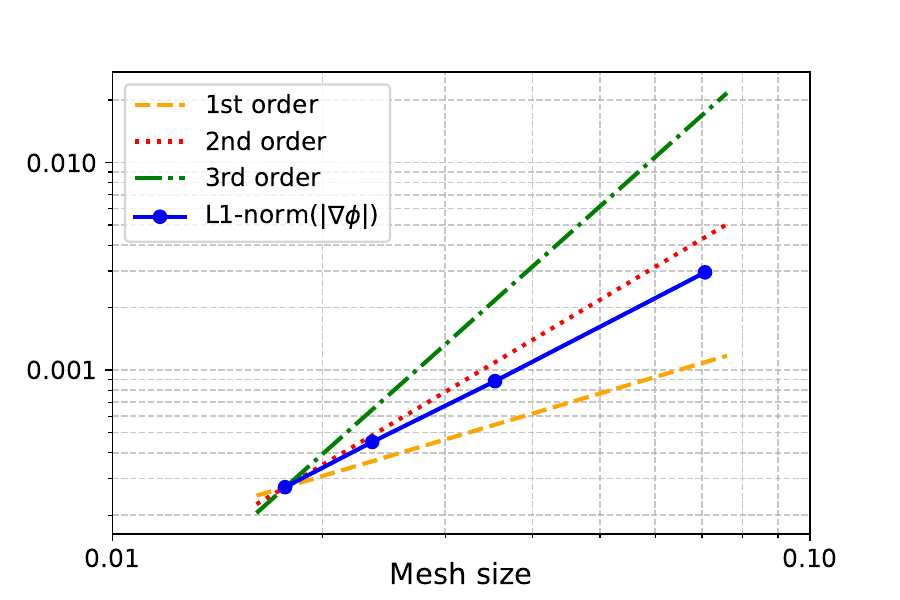}
				\includegraphics[width=0.5\textwidth]{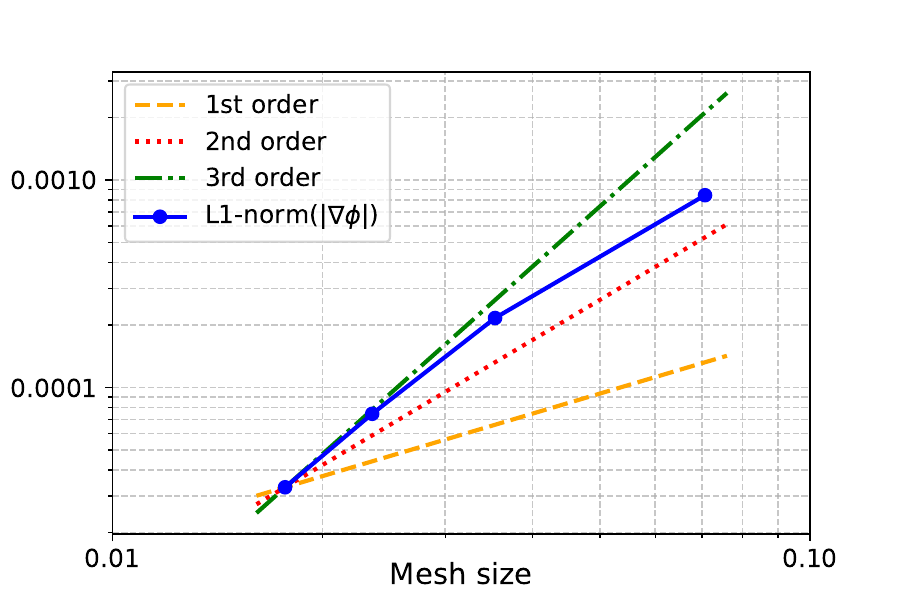}}
		\end{subfigure}
		\caption{Convergence analysis of the $L1$-norm of the error in the gradient of the level-set function by reducing the mesh size ($h$) calculated for the full-band (left column) and the narrow-band (right column). Reference lines depicting first-order (orange dashed), second-order (red dotted), and third-order (green dash-dot) convergence are included to aid visualization. Each row corresponds to the cases with different contact angles depicted in Fig.~\ref{fig:spherical}~a-c, respectively.}
		\label{fig:sphere_convergence}
	\end{figure}
	
	The left column of Fig.~\ref{fig:sphere_convergence} reveals a comprehensive depiction of the convergence behavior of the proposed re-initialization method. This illustration is achieved by systematically reducing the mesh size ($h$) and computing the (average) $L^1$-norm of the error in the gradient of the level-set function for the full-band. The plot demonstrates a convergence pattern that closely aligns with second-order convergence.
	
	The right column of Fig.~\ref{fig:sphere_convergence} provides an analogous analysis, yet the focus shifts to the narrow-band of $-3h < \phi < 3h$. The convergence trends observed here exhibit a remarkable alignment with the third-order convergence line. This observation indicates that the proposed optimization-based re-initialization approach is highly beneficial in terms of accuracy.
	Together, these convergence analyses, depicted in Fig.~\ref{fig:sphere_convergence}, collectively underscore the accuracy and robustness of our proposed method in maintaining the magnitude of the level-set gradient ($|\nabla \phi| = 1$) even in the presence of intricate geometries and dynamic simulations.

	\begin{figure}[!htb]
		\begin{subfigure}{\textwidth}
			\centering
			\makebox[0.5\textwidth][c]{\scriptsize\textbf{interface preservation}}%
			\makebox[0.5\textwidth][c]{\scriptsize\textbf{contact line preservation}}%
			\subcaptionOverlay{\includegraphics[width=0.5\textwidth]{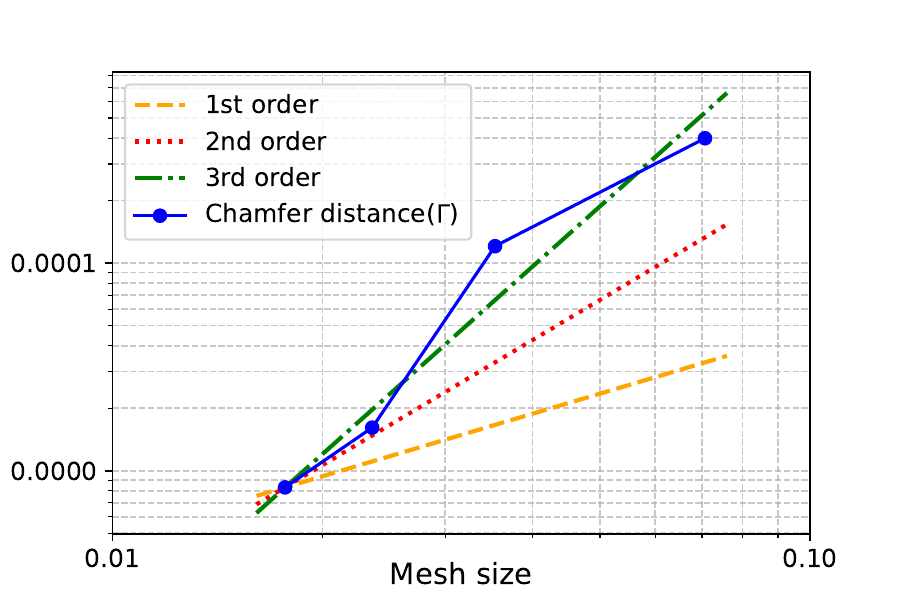}
				\includegraphics[width=0.5\textwidth]{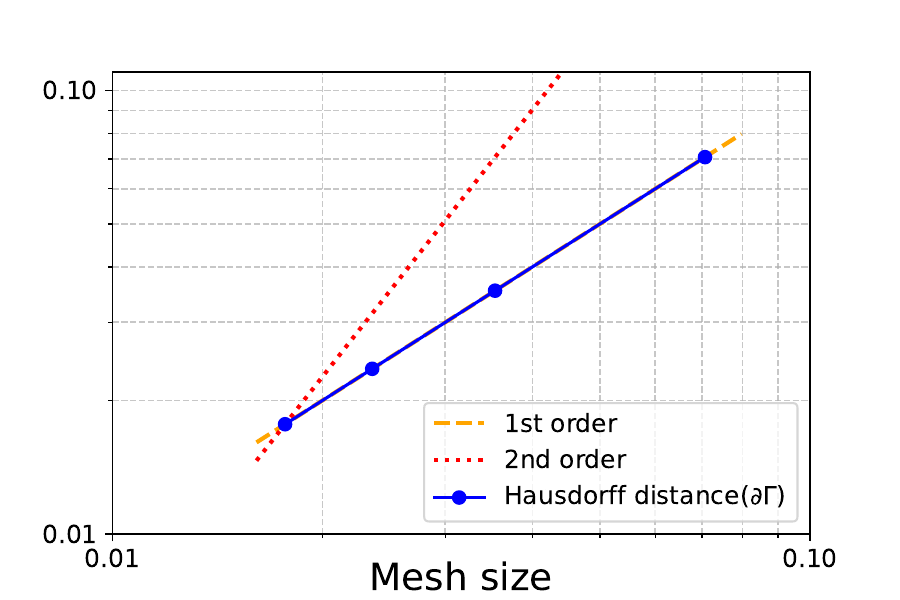}}
		\end{subfigure}
		\begin{subfigure}{\textwidth}
			\centering
			\subcaptionOverlay{\includegraphics[width=0.5\textwidth]{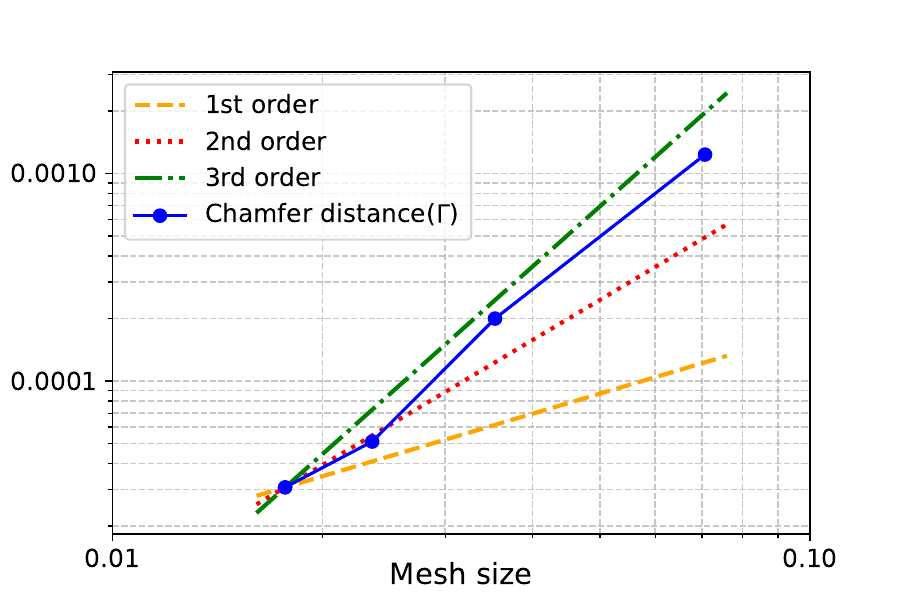}
				\includegraphics[width=0.5\textwidth]{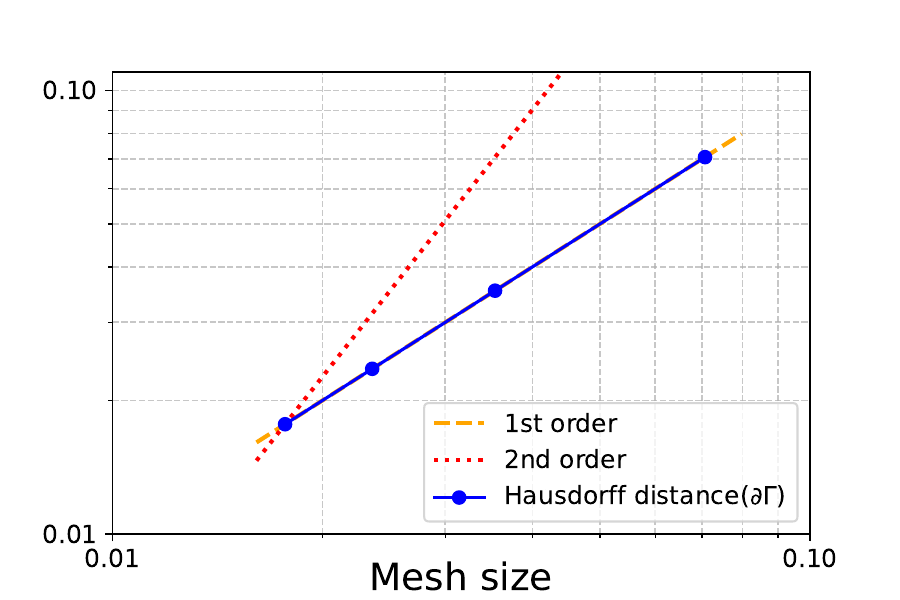}}
		\end{subfigure}
		\begin{subfigure}{\textwidth}
			\centering
			\subcaptionOverlay{\includegraphics[width=0.5\textwidth]{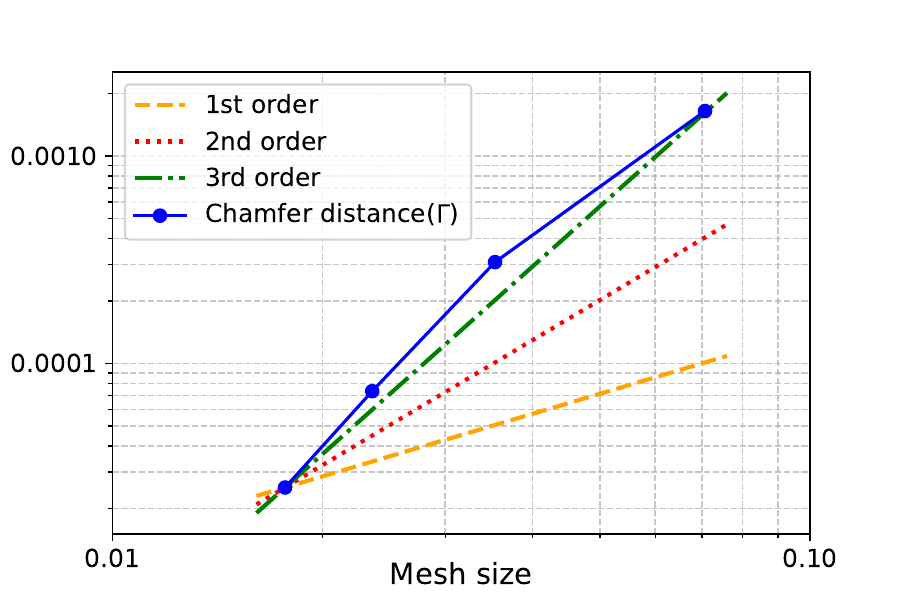}
				\includegraphics[width=0.5\textwidth]{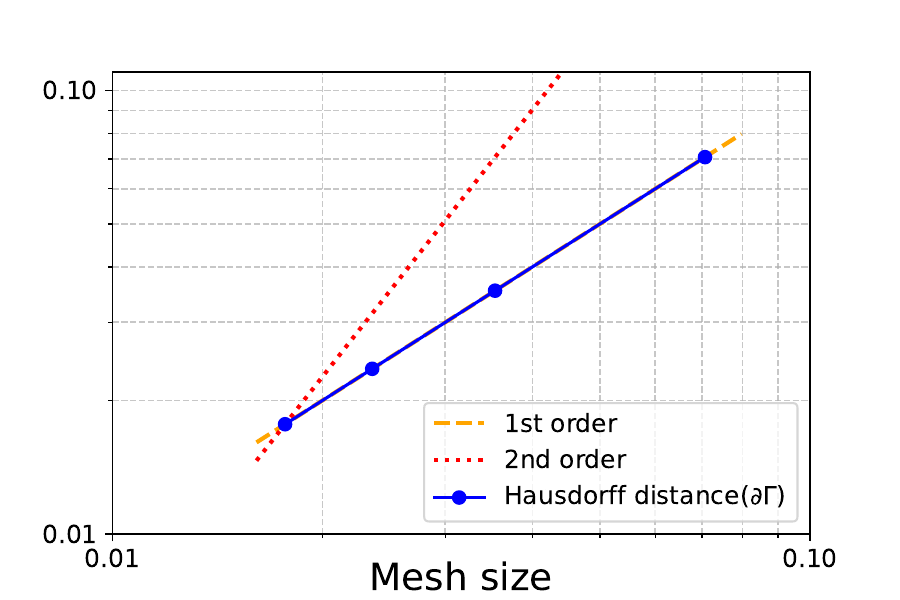}}
		\end{subfigure}
		\caption{Convergence analysis of the Chamfer Distance for the $\phi=0$ iso-surface (left column) and Hausdorff Distance for the contact line (right column) of sessile spherical droplets, achieved by reducing the mesh size ($h$). Reference lines depicting first-order (orange dashed), second-order (red dotted), and third-order (green dash-dot) convergence are included to aid visualization. Each row corresponds to the cases with different contact angles depicted in Fig.~\ref{fig:spherical}~a-c, respectively.}
		\label{fig:spherical_CD}
	\end{figure}
	In Fig.~\ref{fig:spherical_CD} (left column), the Chamfer Distance (Eq.~\ref{eq:CD}) plots for the three spherical droplets with different contact angles are presented.
	The observed third-order convergence behavior indicates the method's accuracy in preserving the interface, highlighting the effectiveness of the element-splitting scheme and optimization approach in our method. 
	
	Similarly, the right column of Fig.~\ref{fig:spherical_CD} displays a first-order convergence trend in Hausdorff Distance (Eq.~\ref{eq:HD}), showcasing the method's capability in accurately preserving the contact line even in relatively coarser meshes.
	
	These figures collectively underscore the effectiveness and accuracy of the proposed level-set re-initialization method in preserving both the interface and contact line for spherical interfaces.

	\subsubsection{Ellipsoidal droplet}
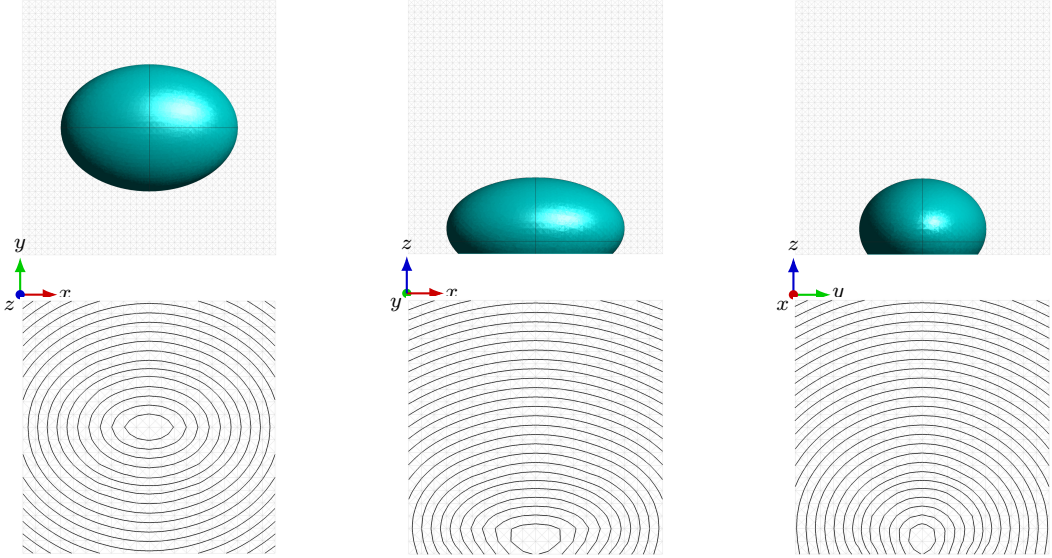
\begin{figure}[!htb]
		\centering
		\begin{minipage}{.25\textwidth}
			\begin{subfigure}{\textwidth}      
				\includegraphics[width=\textwidth]{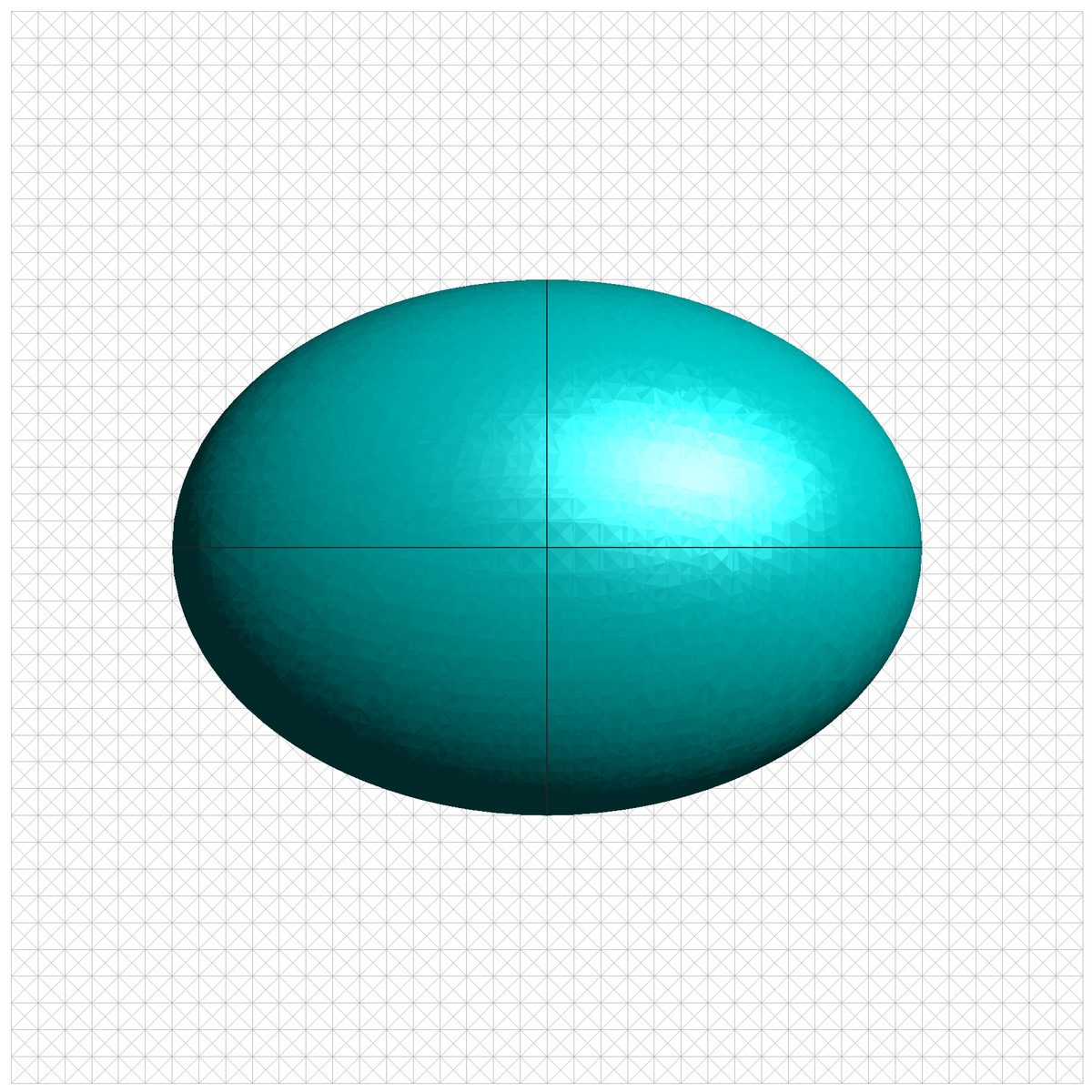}
				\begin{tikzpicture}[overlay,remember picture,scale=0.5]
					\draw[-{Latex[length=2mm]}, red!80!black, thick] (0,0,0) -- (1,0,0);
					\draw[-{Latex[length=2mm]}, green!80!black, thick] (0,0,0) -- (0,1,0);  
					\fill[blue!80!black] (0,0,0) circle[radius=3.5pt];
					\node at (1.2,0,0) {\scriptsize$x$};
					\node at (0,1.3,0) {\scriptsize$y$};
					\node at (0.1,0.1,1) {\scriptsize$z$};
				\end{tikzpicture}
			\end{subfigure}
			\begin{subfigure}{\textwidth}            
				\includegraphics[width=\textwidth]{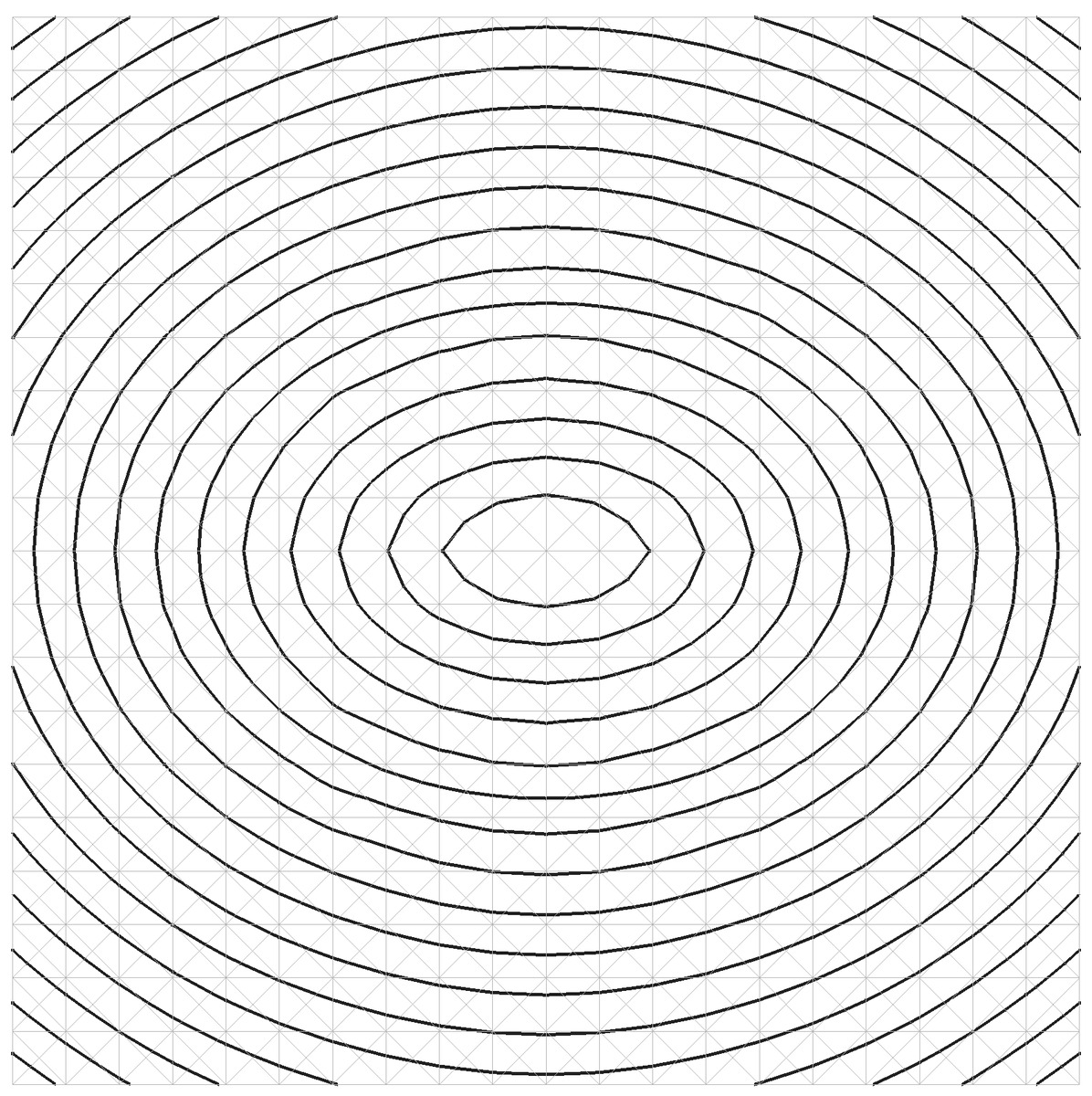}
			\end{subfigure}    
		\end{minipage}
		\hfill
		\begin{minipage}{.25\textwidth}
			\begin{subfigure}{\textwidth}      
				\includegraphics[width=\textwidth]{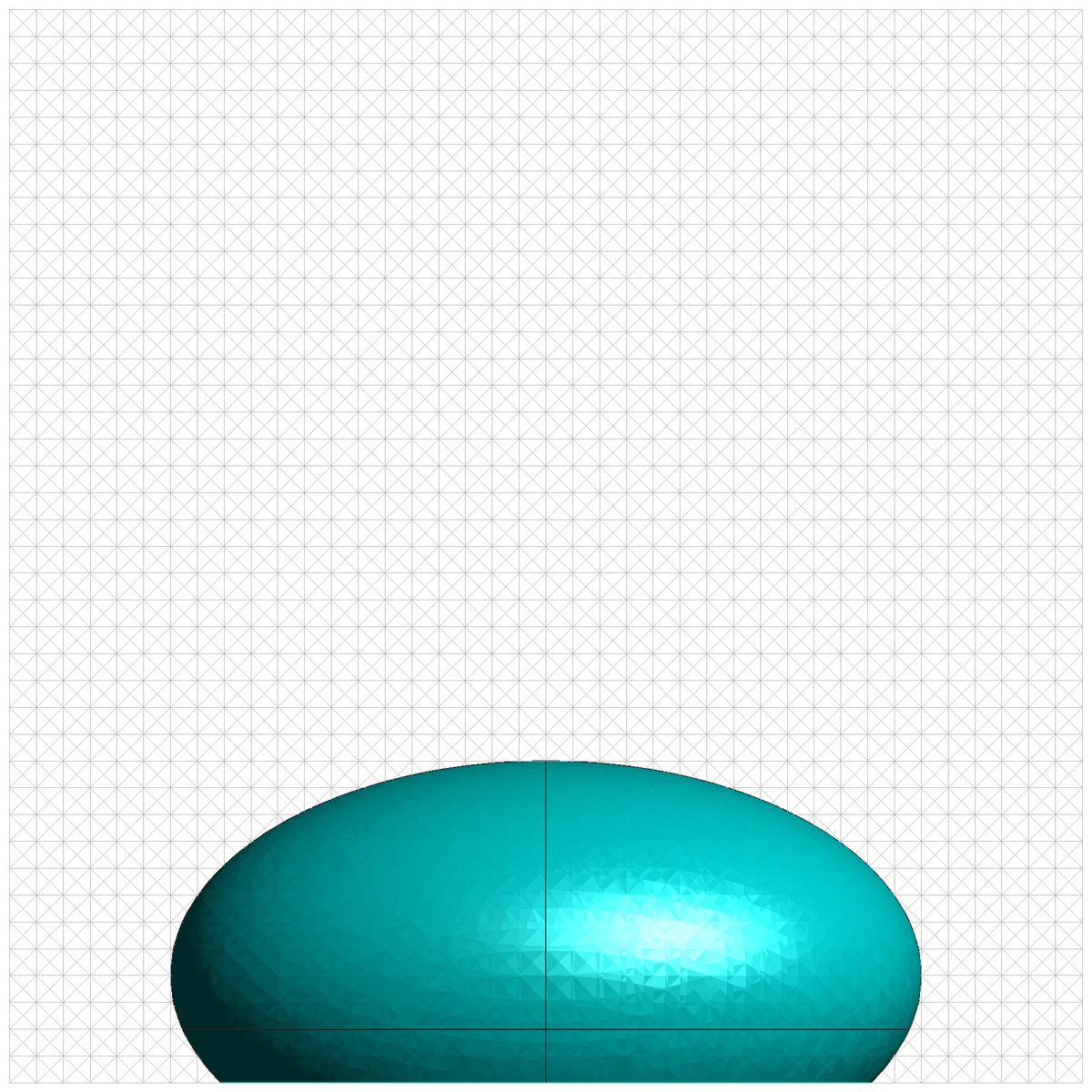}
				\begin{tikzpicture}[overlay,remember picture,scale=0.5]
					\fill[green!80!black] (0,0,0) circle[radius=3.5pt];
					\draw[-{Latex[length=2mm]}, red!80!black, thick] (0,0,0) -- (1,0,0);
					\draw[-{Latex[length=2mm]}, blue!80!black, thick] (0,0,0) -- (0,1,0);
					\node at (1.2,0,0) {\scriptsize$x$};
					\node at (0,1.3,0) {\scriptsize$z$};
					\node at (0.1,0.1,1) {\scriptsize$y$};
				\end{tikzpicture}
			\end{subfigure}
			\begin{subfigure}{\textwidth}      
				\includegraphics[width=\textwidth]{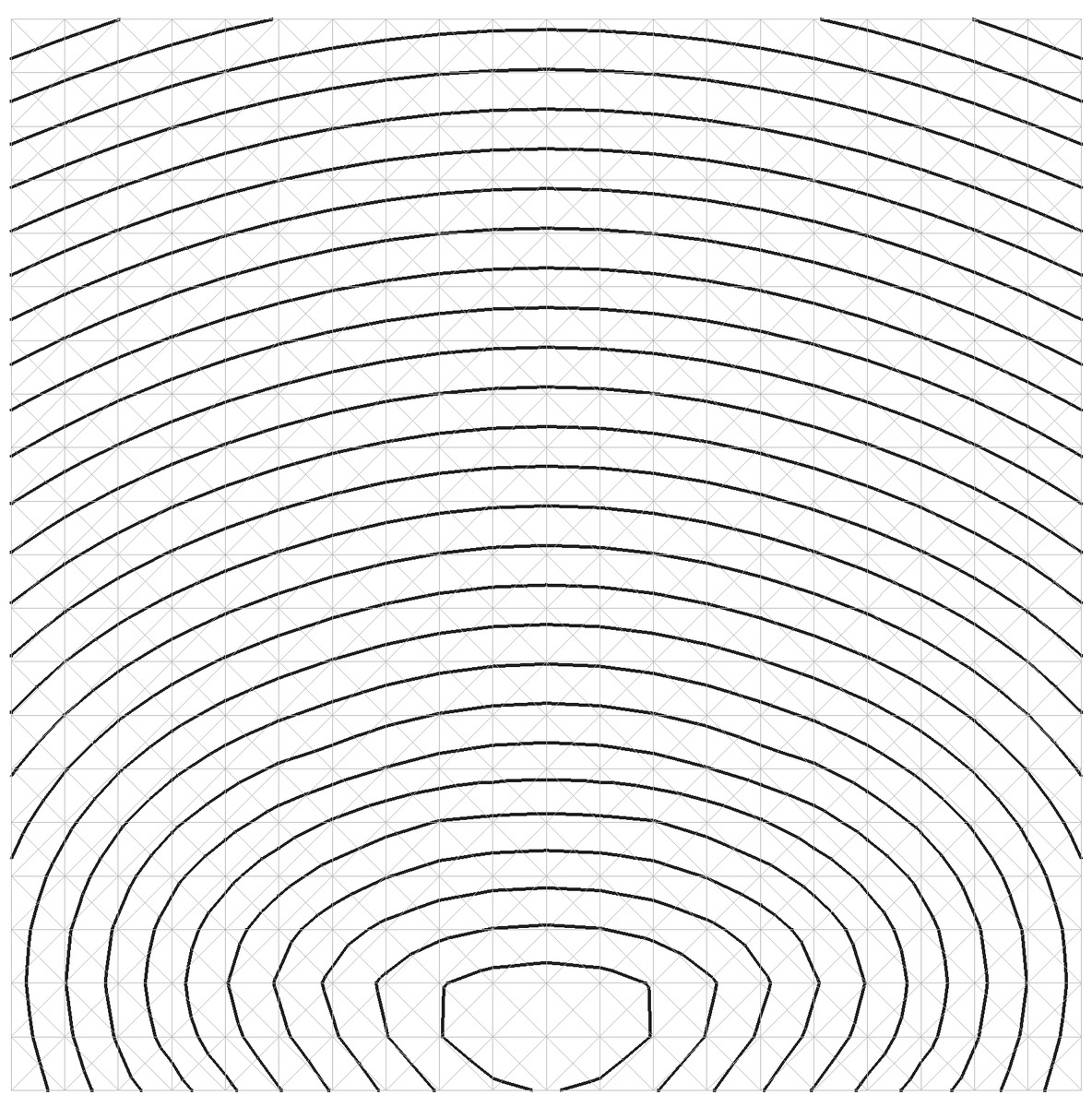}
			\end{subfigure}    
		\end{minipage}
		\hfill
		\begin{minipage}{.25\textwidth}
			\begin{subfigure}{\textwidth}     
				\includegraphics[width=\textwidth]{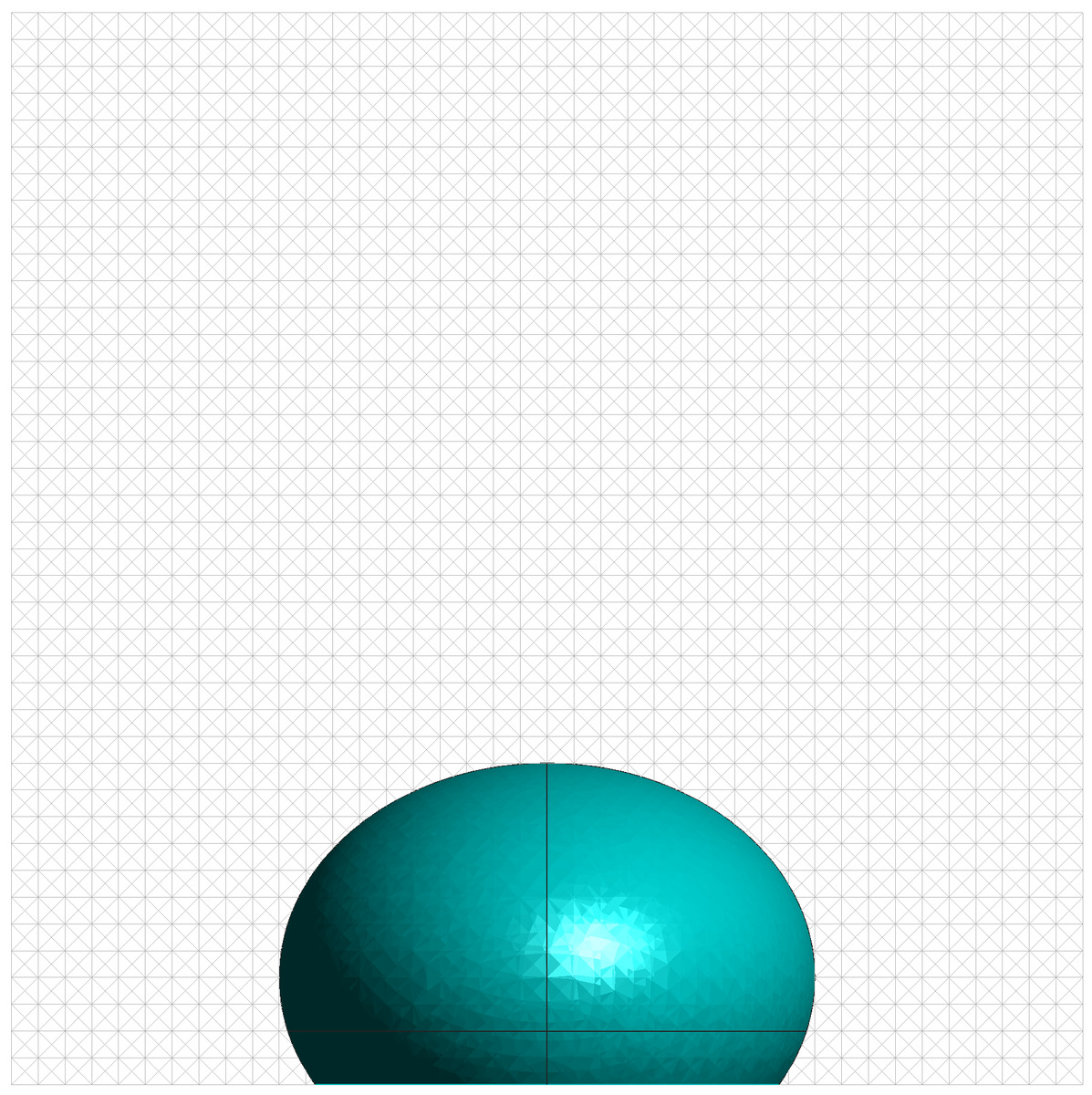}
				\begin{tikzpicture}[overlay,remember picture,scale=0.5]
					\draw[-{Latex[length=2mm]}, green!80!black, thick] (0,0,0) -- (1,0,0);
					\draw[-{Latex[length=2mm]}, blue!80!black, thick] (0,0,0) -- (0,1,0);  
					\fill[red!80!black] (0,0,0) circle[radius=3.5pt];
					\node at (1.2,0,0) {\scriptsize$y$};
					\node at (0,1.3,0) {\scriptsize$z$};
					\node at (0.1,0.1,1) {\scriptsize$x$};
				\end{tikzpicture}
			\end{subfigure}
			\begin{subfigure}{\textwidth}        
				\includegraphics[width=\textwidth]{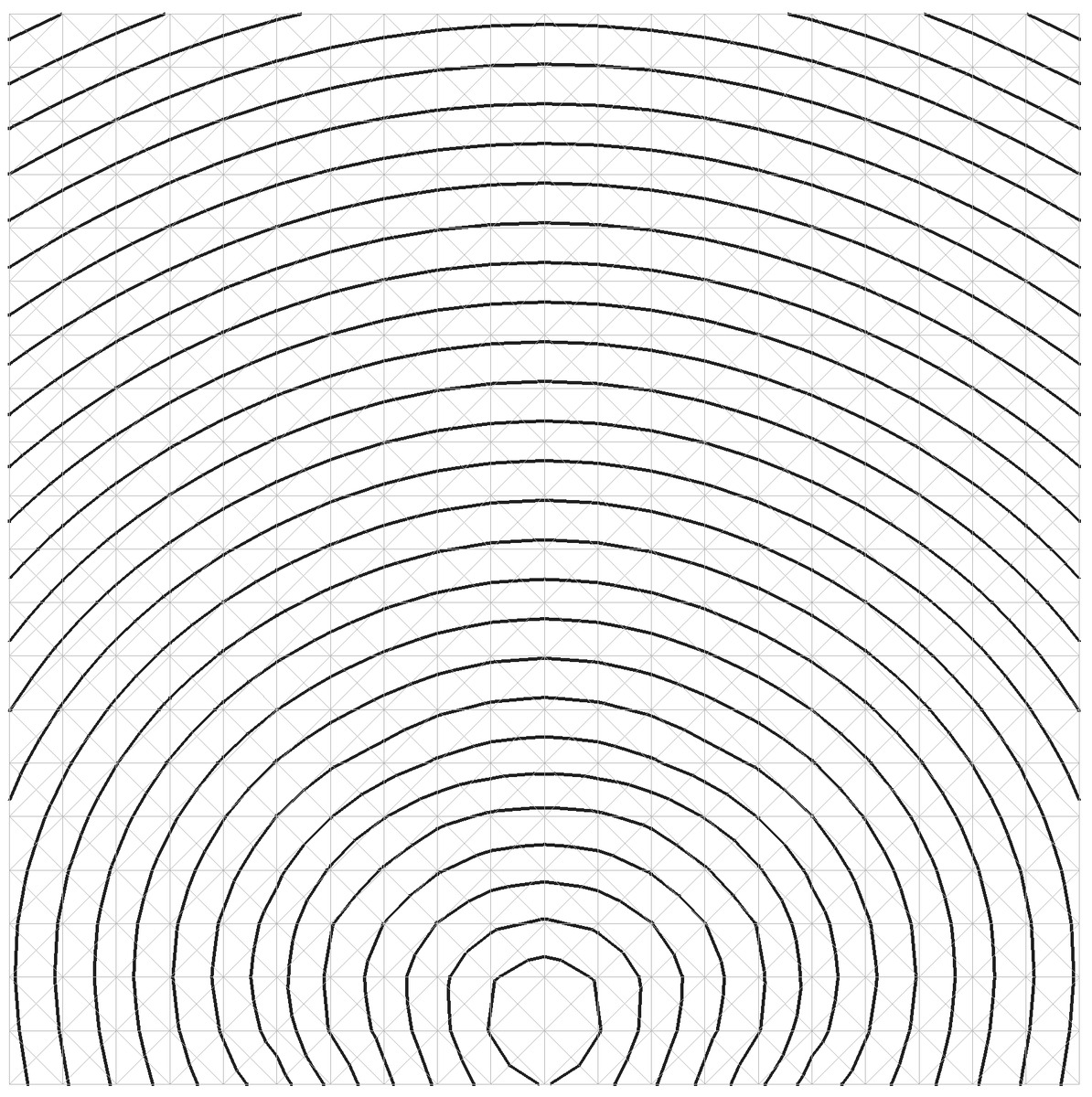}
			\end{subfigure}    
		\end{minipage}
		\caption{Side views of a sessile ellipsoidal droplet from three different perspectives: the $xy$, $xz$, and $yz$ planes (upper row), as well as, the corresponding contours of iso-level-set (lower row)}
		\label{fig:ellipsoid}
	\end{figure}

	Building upon the spherical droplet case, we introduce a little added complexity by simulating ellipsoidal droplets with three semi-axes lengths of $a=350 \mu$m, $b=250 \mu$m, and $c=200 \mu$m along the $x$, $y$, and $z$ axes, respectively. This allows us to demonstrate the generalizability of the proposed method to handle non-axisymmetric shapes. In Fig.~\ref{fig:ellipsoid}, side views of the ellipsoidal droplet from the $xy$, $xz$, and $yz$ planes are presented, providing a comprehensive visualization of its geometry. The lower row displays iso-level-set contours for each side view, demonstrating the regularity of the reinitialized level-set.
	
\begin{figure}[!htb]
		\begin{subfigure}{\textwidth}
			\centering
			\makebox[0.5\textwidth][c]{\scriptsize\textbf{full-band}}%
			\makebox[0.5\textwidth][c]{\scriptsize\textbf{narrow-band}}%
			\subcaptionOverlay{\includegraphics[width=0.5\textwidth]{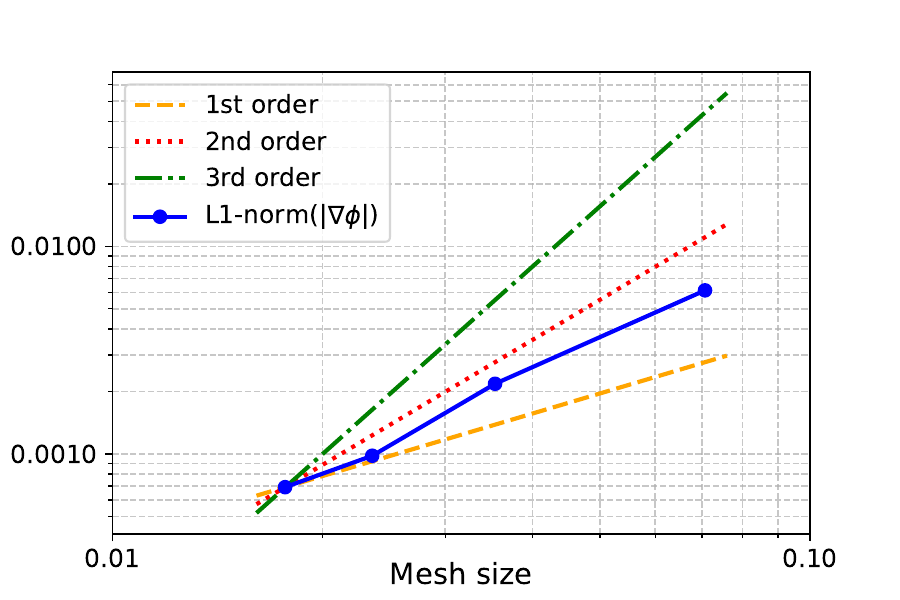}
				\includegraphics[width=0.5\textwidth]{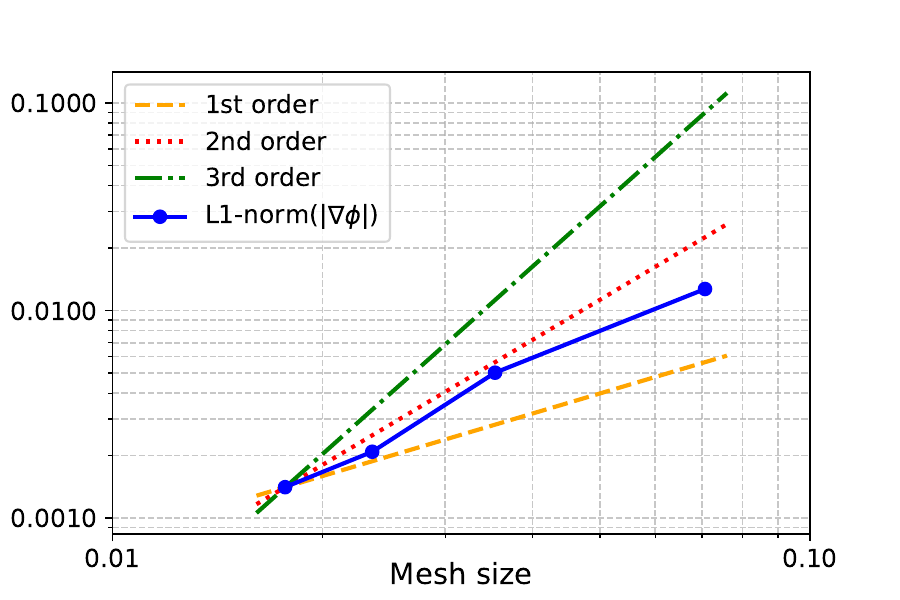}}
		\end{subfigure}    
		\hfill  
		\caption{Convergence analysis of the $L^1$-norm error in the gradient of the level-set function of sessile ellipsoidal droplets by reducing the mesh size ($h$), computed over the full-band (left column) and the (right column). Reference lines depicting first-order (orange dashed), second-order (red dotted), and third-order (green dash-dot) convergence are included for clarity. 
		}
		\label{fig:ellipsoid_convergence}
	\end{figure}

\begin{figure}[!htb]
		\begin{subfigure}{\textwidth}
			\centering
			\makebox[0.5\textwidth][c]{\scriptsize\textbf{interface preservation}}%
			\makebox[0.5\textwidth][c]{\scriptsize\textbf{contact line preservation}}%
			\subcaptionOverlay{\includegraphics[width=0.5\textwidth]{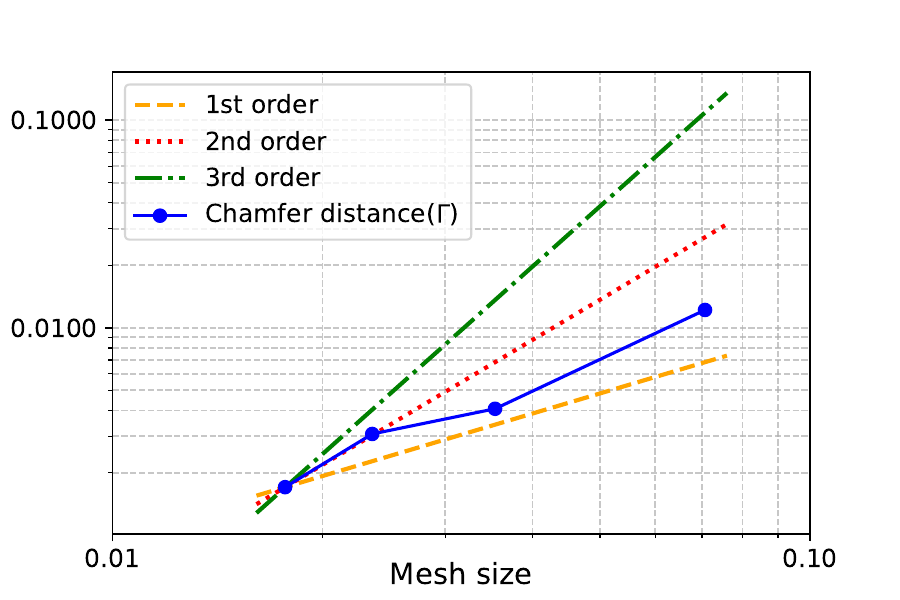}
				\includegraphics[width=0.5\textwidth]{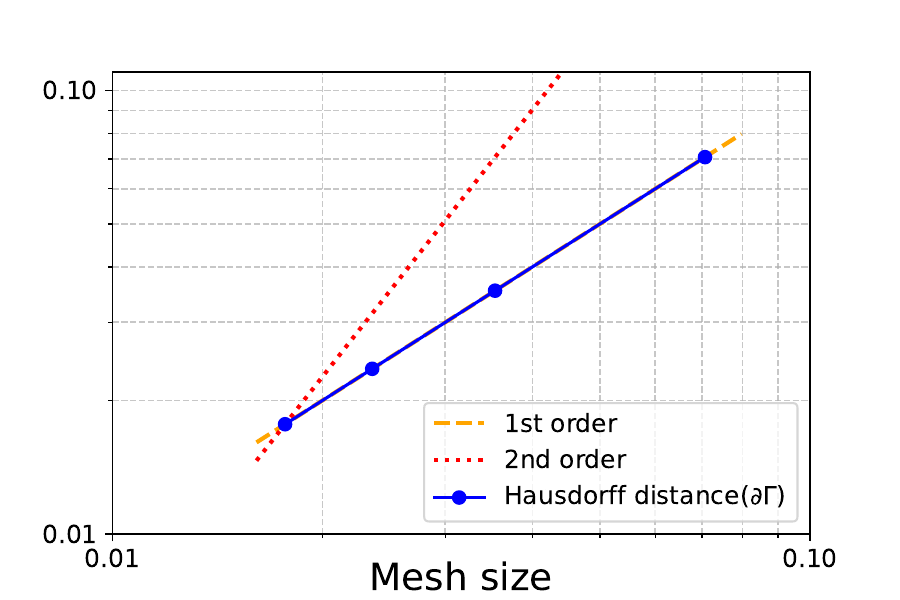}}
		\end{subfigure}  
		\caption{Convergence analysis of the Chamfer Distance for the $\phi=0$ iso-surface (left column) and Hausdorff Distance for the contact line (right column) of the sessile ellipsoidal droplet, achieved by reducing the mesh size ($h$). Reference lines depicting first-order (orange dashed), second-order (red dotted), and third-order (green dash-dot) convergence are included to aid visualization.
		}
		\label{fig:ellipsoid_CD}
	\end{figure}

	Fig.~\ref{fig:ellipsoid_convergence} demonstrates second-order convergence for the $L^1$-norm of $|\nabla \phi|$ in both the full-band and narrow-band cases of the ellipsoidal droplet. 
	Furthermore, as illustrated in Fig.~\ref{fig:ellipsoid_CD}, the proposed method showcases second-order convergence for accurately preserving the ellipsoidal liquid-gas interface ($\phi=0$ iso-surface), while maintaining first-order convergence in preserving the contact line.
	
	\subsection{Dynamic Test Cases}
	\subsubsection{Oscillating Free Droplet}\label{sec:OsciDrop}
\begin{figure}[!htb]
		\centering
		\includegraphics[width=0.32\textwidth]{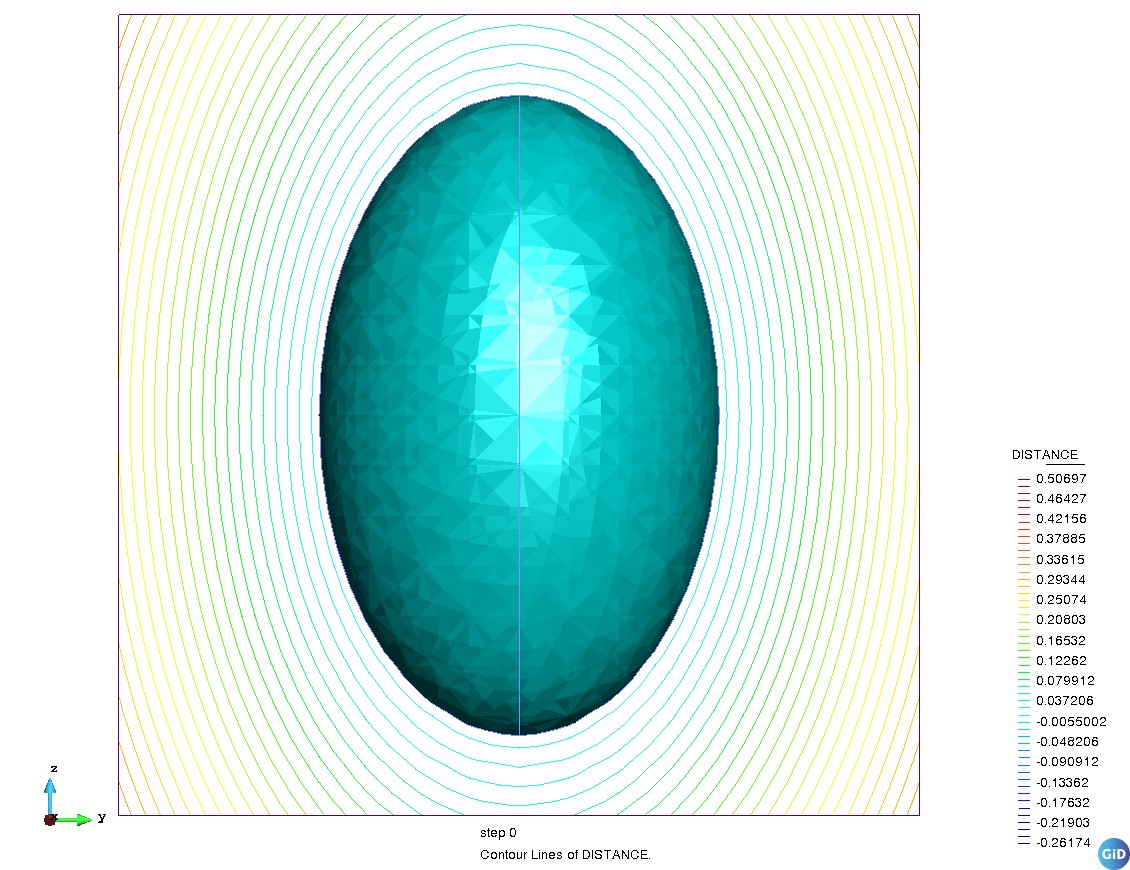}
		\includegraphics[width=0.32\textwidth]{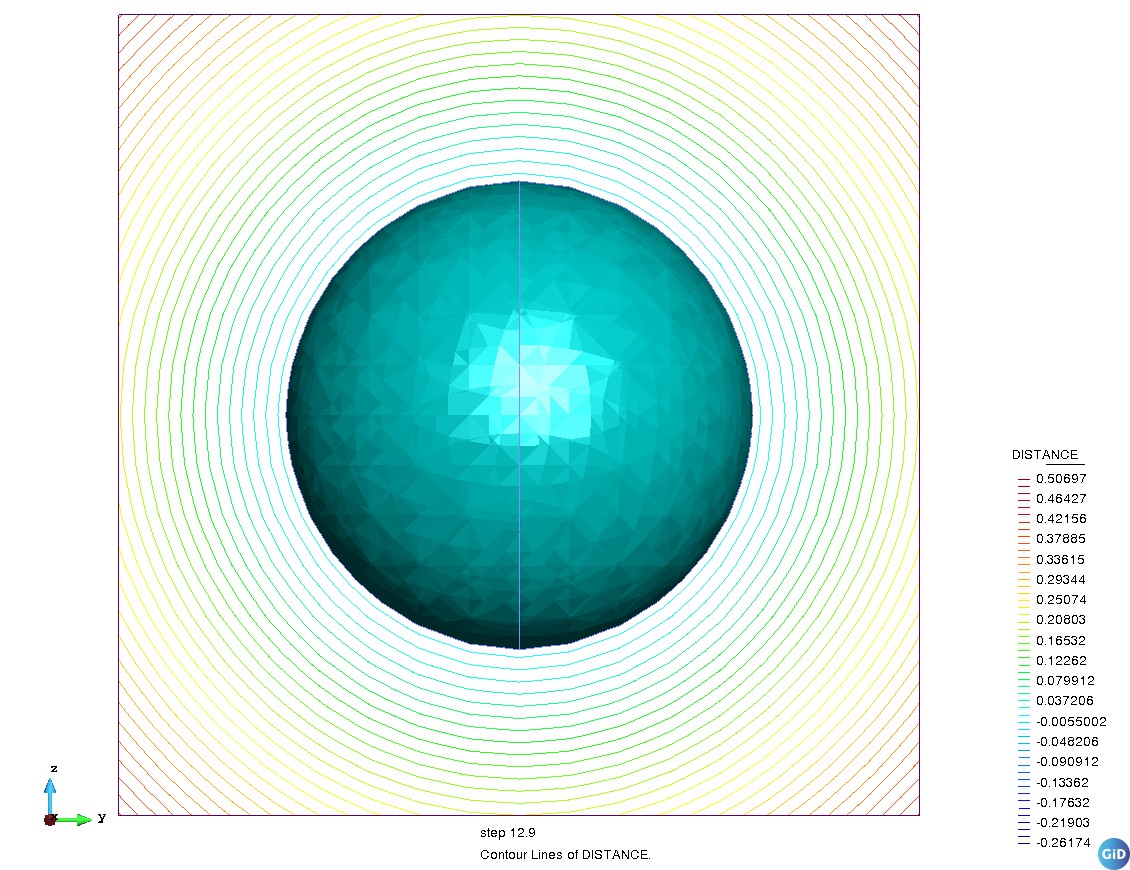}
		\includegraphics[width=0.32\textwidth]{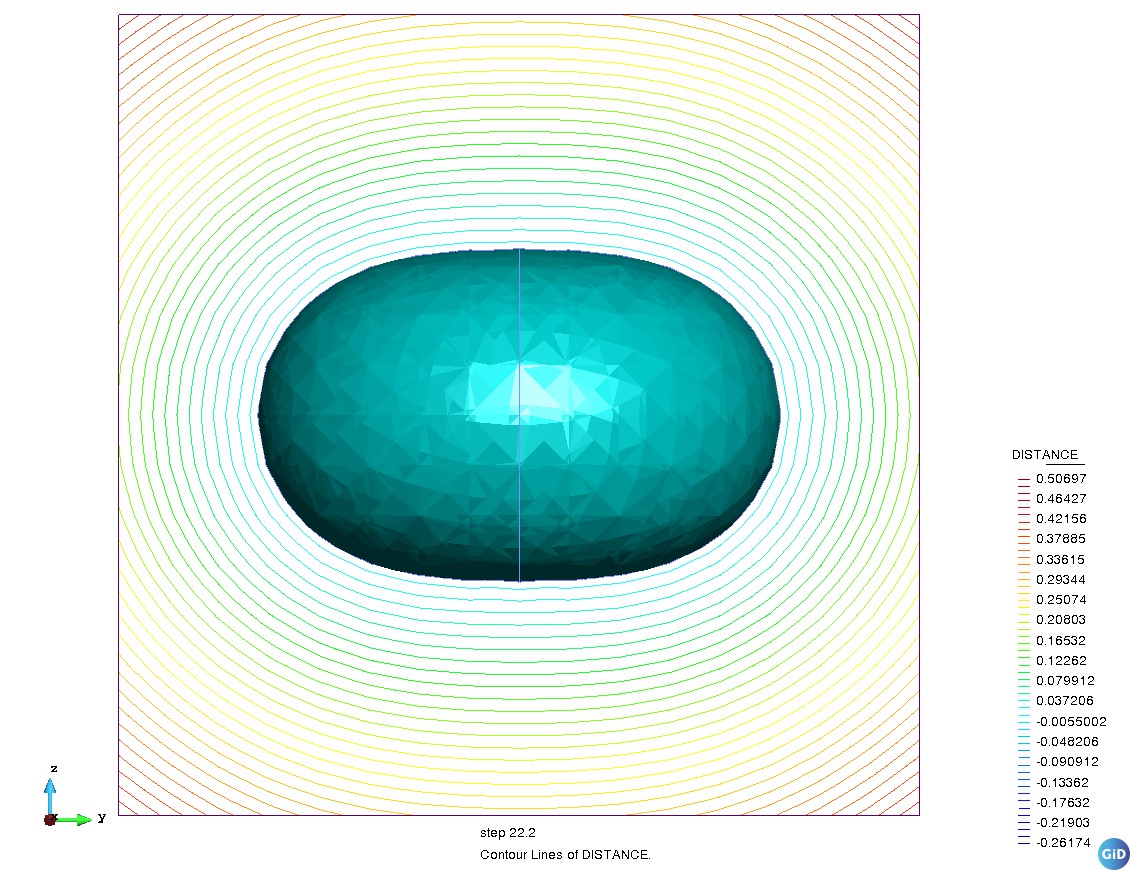}
		\caption{Temporal evolution of an oscillating droplet}
		\label{fig:oscillating}
	\end{figure}

	This dynamic scenario explores the evolution of the droplet's interface, driven solely by surface tension forces. This test case introduces additional complexity to the simulation, as the surface tension forces cause the droplet to deform and change shape over time (Fig.~\ref{fig:oscillating}). It allows us to examine the method's capability to preserve the zero level-set surface accurately while the interface dynamically evolves during the simulation.
	The initial geometry of the ellipsoidal droplet is defined with semi-axes of $a=2.5\mu$m, $b=2.5\mu$m, and $c=4.0\mu$m along the $x$, $y$, and $z$ axes, respectively. To emphasize the influence of surface tension, the gravitational force is intentionally excluded from the simulation.
	
	\begin{figure}[!htb]    
		\begin{subfigure}{0.5\textwidth}\centering
			\subcaptionOverlay{\includegraphics[width=\textwidth]{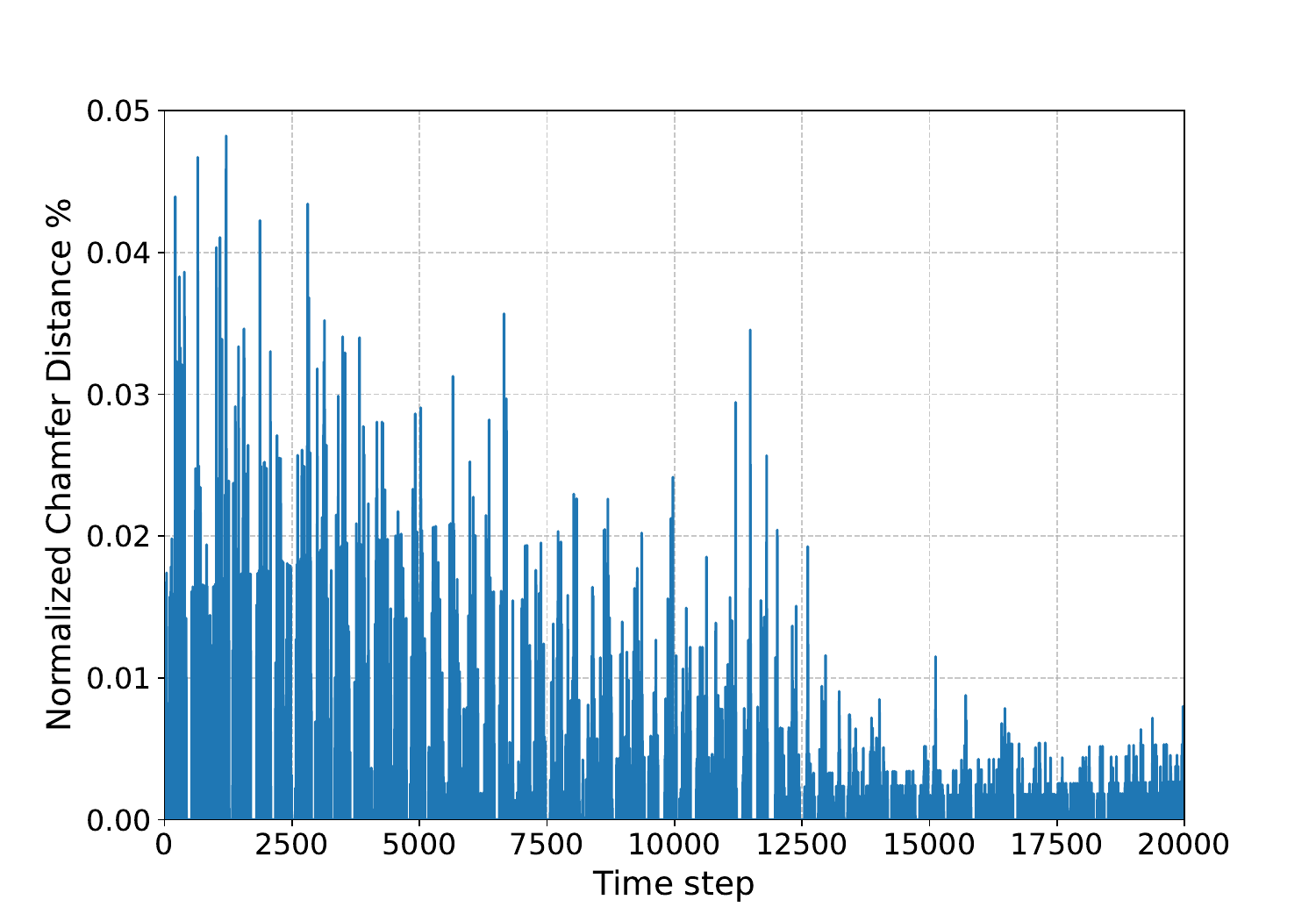}}
		\end{subfigure}
		\begin{subfigure}{0.5\textwidth}\centering
			\subcaptionOverlay{\includegraphics[width=\textwidth]{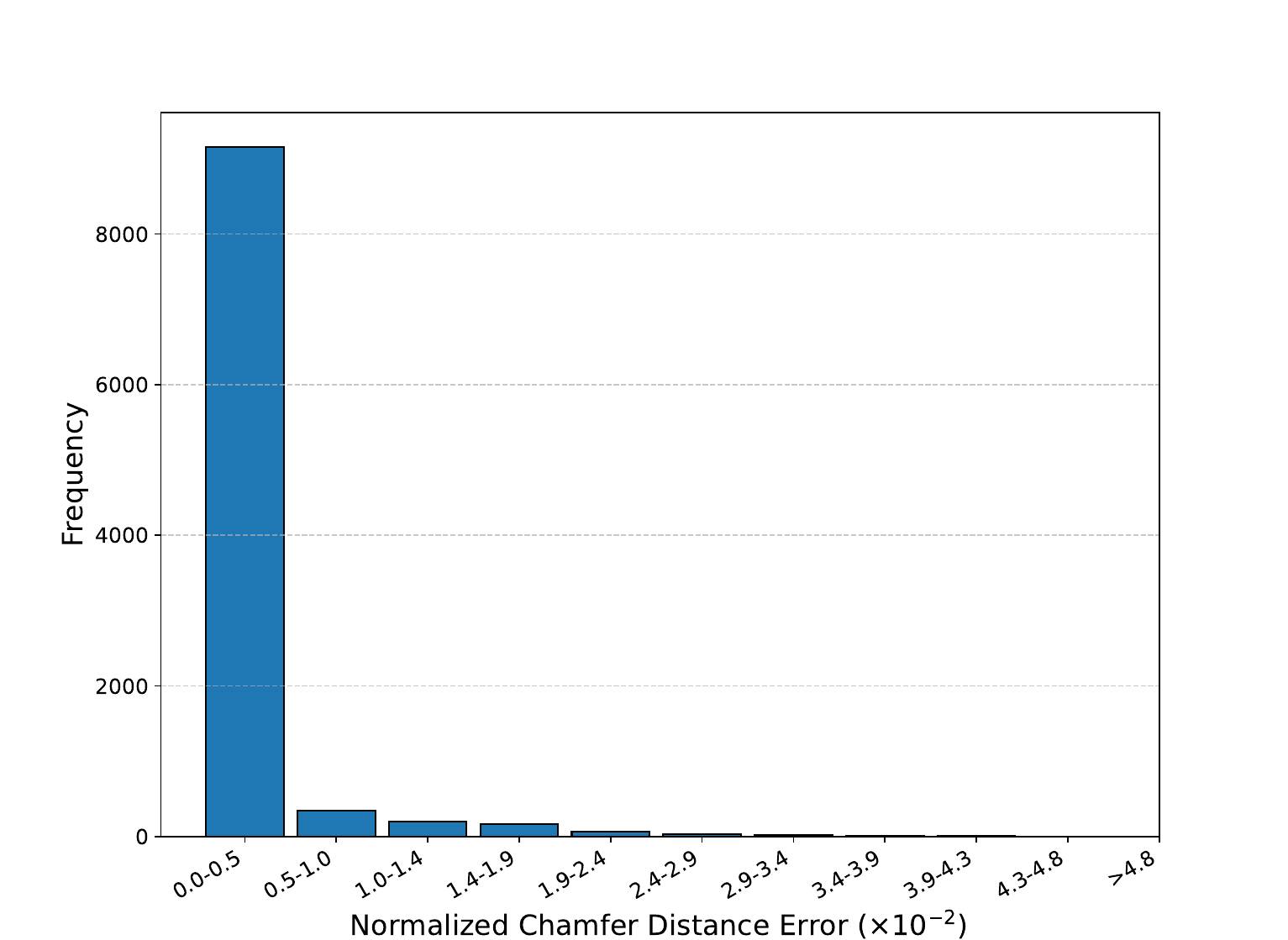}}
		\end{subfigure}
		\caption{a) Time evolution of the Chamfer distance error for the oscillating droplet. The error is normalized to the mean radius of the droplet. b) Distribution of the normalized Chamfer distance error during simulation time steps}
		\label{fig:oscilating_error}
	\end{figure}
	Fig.~\ref{fig:oscilating_error}(a) illustrates the time evolution of the Chamfer distance error for the oscillating droplet, with the error values normalized to the mean radius of the droplet. Remarkably low error levels reveal our method's proficiency in preserving the $\phi=0$ iso-surface. Notably, as time elapses, the error diminishes, reflecting the transformation of the droplet's shape from ellipsoid to sphere during oscillations. This result aligns with our earlier findings, which showcased the method's superior performance in preserving spherical interfaces compared to ellipsoidal ones.
	
	In Fig.~\ref{fig:oscilating_error}(b), the distribution of normalized Chamfer distance errors across various time steps is depicted in the form of a histogram bar plot. While the maximum normalized error remains comfortably below 0.05 per cent, an important observation emerges—the first bar, representing the range from 0 to 0.005 per cent, is notably larger than the rest. This observation underscores the robustness of our approach in consistently preserving the interface throughout dynamic simulations, particularly during the level-set re-initialization process.

	\subsubsection{Droplet in Channel}\label{sec:DropletChannel}
	
	The second subsection presents the simulation results of a challenging test case involving a droplet confined in a channel. The droplet's interaction with the channel walls and the effects of flow conditions make this test case particularly demanding in terms of accurately capturing the interface evolution.
	Confined droplets exposed to airflow are found in many important practical applications, including the flow channels of fuel cells, where complex droplet-airflow interactions are crucial for the correct evacuation of residual liquid from the device, an essential aspect of its reliable functionality \cite{hashemi_toward_2021}. Additionally, this phenomenon has relevance in various other areas such as drug delivery, material synthesis, and lab-on-chip devices, where micro-scale droplet behavior in confined spaces plays a pivotal role~\cite{dangla_droplet_2013,liu_microfluidics_2021,giannitelli_droplet_based_2022}.
	
\begin{figure}[!htb]
		\begin{subfigure}{\textwidth}
			\centering
			\subcaptionOverlay{\includegraphics[width=0.85\textwidth]{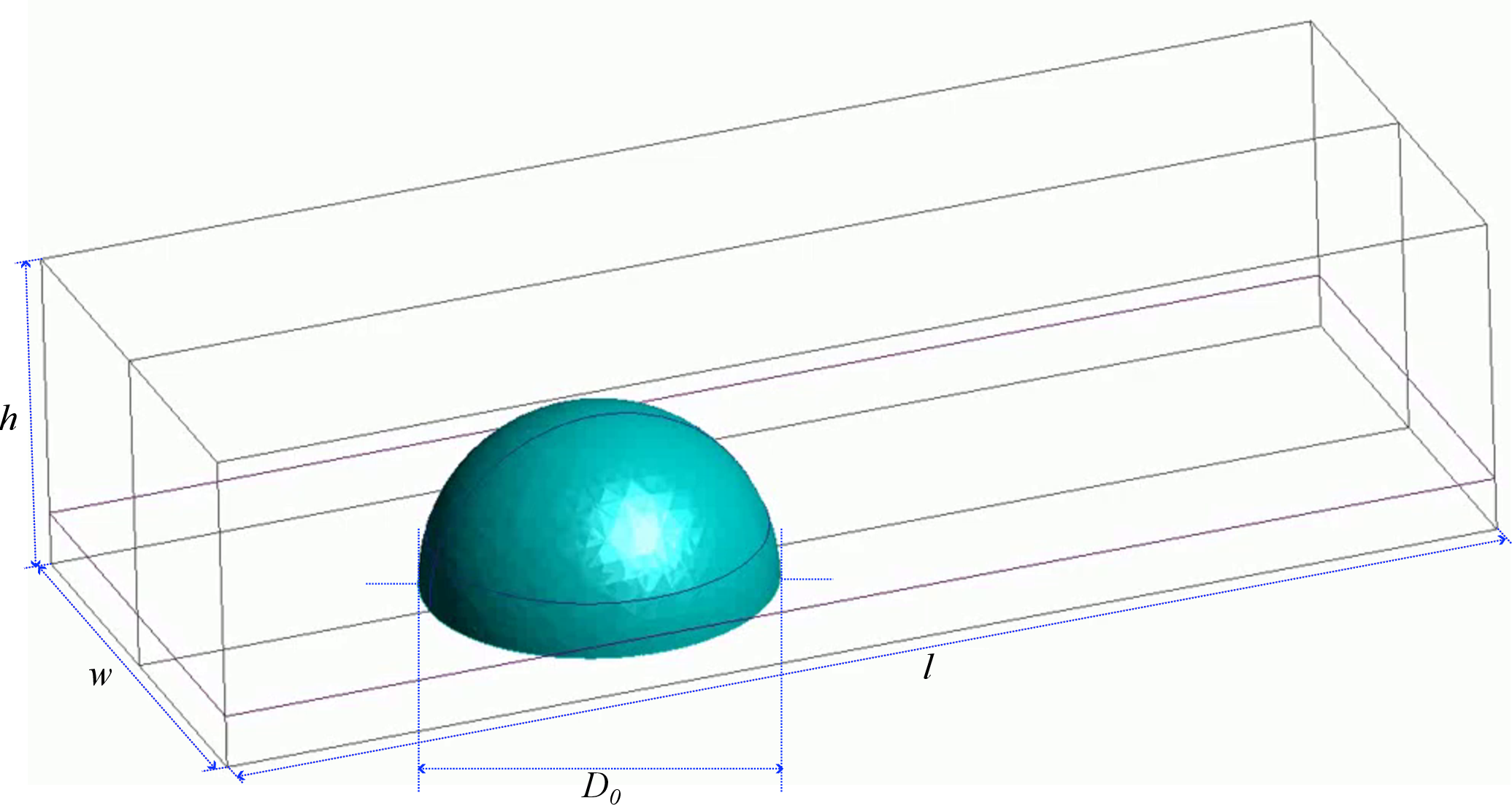}}
		\end{subfigure}
		\begin{subfigure}{\textwidth}
			\centering
			\subcaptionOverlay{\includegraphics[width=0.99\textwidth]{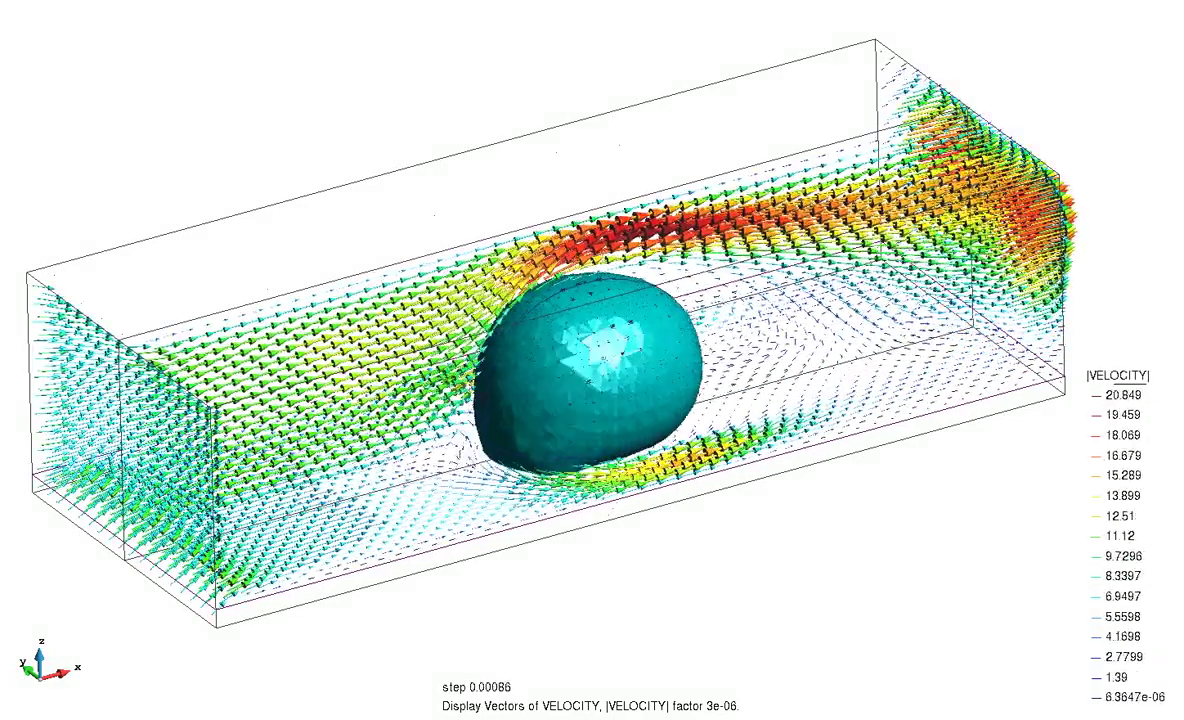}}
		\end{subfigure}    
		\caption{a) The geometry of the test case. b) Snapshot of the 3D view of the simulation results of droplet influenced by the airflow within the channel.}
		\label{fig:channel}
	\end{figure}

	The geometry of the test case, as illustrated in Fig.~\ref{fig:channel}(a), includes a channel with a height of $h=200 \mu$m, a width of $w=300 \mu$m, and a length of $l=800 \mu$m. The initial diameter and the contact angle of the droplet are set to $D_0=214 \mu$m, and $90$ degrees, respectively. For the inlet boundary condition, a prescribed velocity profile is applied in the $x$-direction, described by the following equation:
	\begin{equation}
		u(t) =
		\begin{cases}
			\frac{u_0}{2}\left(1-\cos\left(\frac{\pi}{0.001}t\right)\right) & \text{if } t \leq 0.001\text{s}, \\
			u_0 & \text{if } t > 0.001\text{s}.
		\end{cases}
	\end{equation}
	At the outlet, a constant pressure boundary condition (zero pressure) is enforced. Additionally, the initial position of the droplet is set to be $1.5h$ away from the inlet, aiming to minimize the influence of the spatially uniform velocity applied at the domain's boundary. The time step is chosen as $\Delta t = 10^{-6}$ s, and the computational domain is discretized into approximately 250,000 elements.
	
	Fig.~\ref{fig:channel}(b) provides a snapshot of the 3D view of the simulation results, capturing the moment when the droplet is influenced by the airflow within the channel. We conducted two simulations for this scenario. In the first simulation, the Reynolds number was set to 140 ($u_0\approx 5$ m/s). Confident in our method's capability to handle higher Reynolds numbers effectively, we performed a second simulation with intensified airflow to reach a Reynolds number of 200 ($u_0\approx 7$ m/s). This increase in airflow intensity eventually causes the droplet to detach from the substrate and begin moving within the channel. For a comprehensive view of the complete temporal evolution of these simulations, readers may refer to the  animation (see the video uploaded as supplementary material).
	
	\begin{figure}[!htb]
		\centering
		\begin{minipage}{.30\textwidth}
			\begin{subfigure}{\textwidth}
				\includegraphics[width=\textwidth]{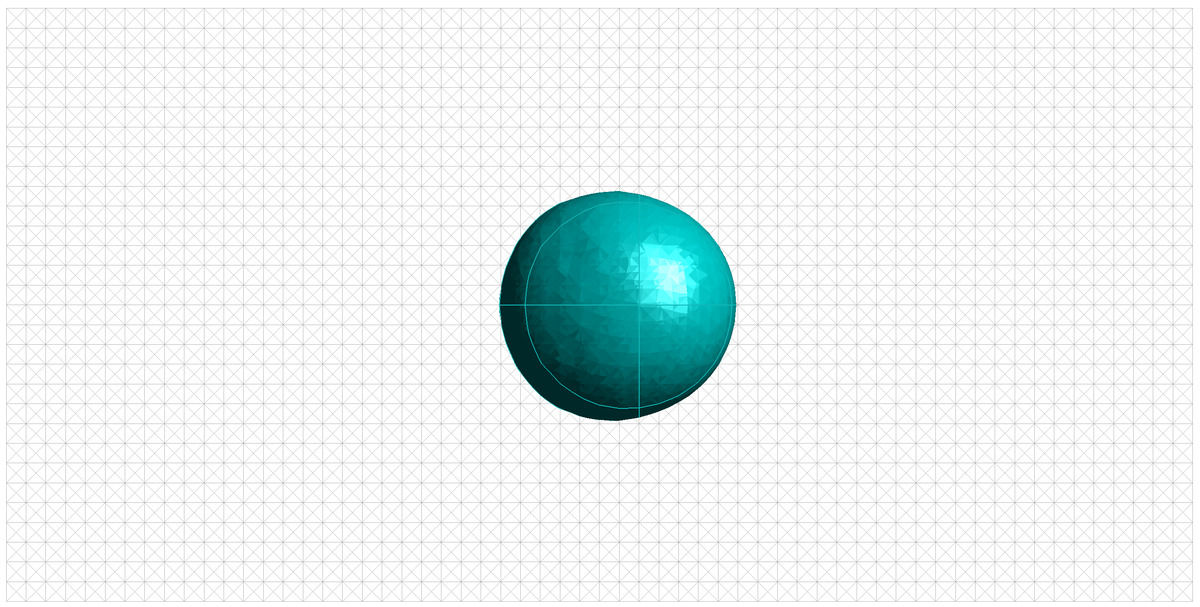}
				\begin{tikzpicture}[overlay,remember picture,scale=0.5]
					\draw[-{Latex[length=2mm]}, red!80!black, thick] (0,0,0) -- (1,0,0);
					\draw[-{Latex[length=2mm]}, green!80!black, thick] (0,0,0) -- (0,1,0);  
					\fill[blue!80!black] (0,0,0) circle[radius=3.5pt];
					\node at (1.2,0,0) {\scriptsize$x$};
					\node at (0,1.3,0) {\scriptsize$y$};
					\node at (0.1,0.1,1) {\scriptsize$z$};
				\end{tikzpicture}
			\end{subfigure}
			\begin{subfigure}{\textwidth}            
				\includegraphics[width=\textwidth]{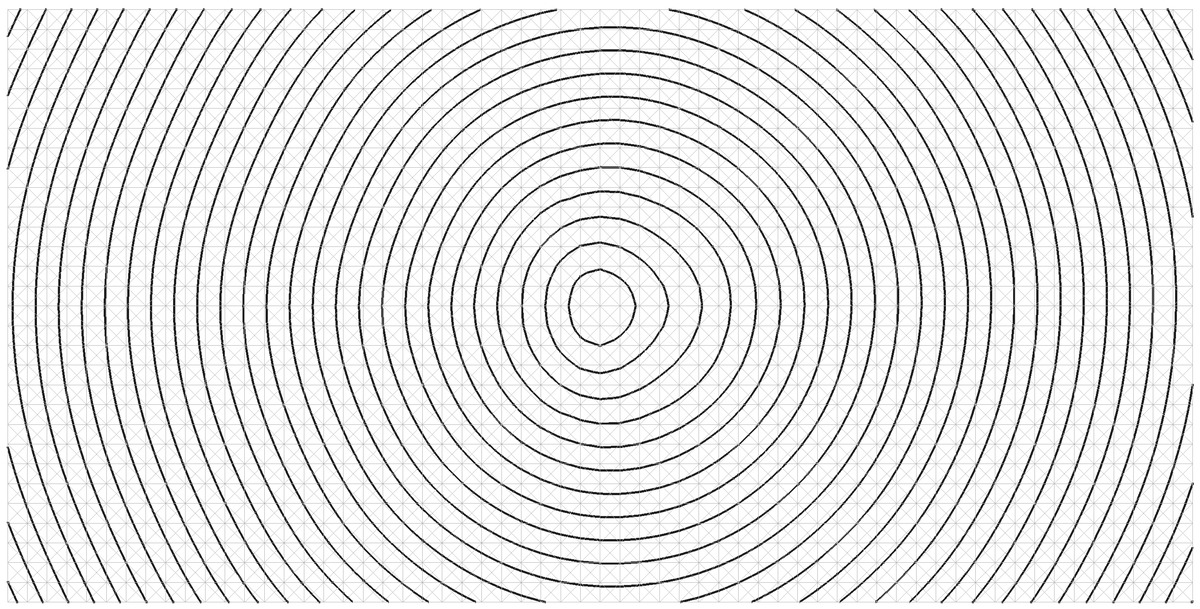}
			\end{subfigure}    
		\end{minipage}
		\hfill
		\begin{minipage}{.37\textwidth}
			\begin{subfigure}{\textwidth}      
				\includegraphics[width=\textwidth]{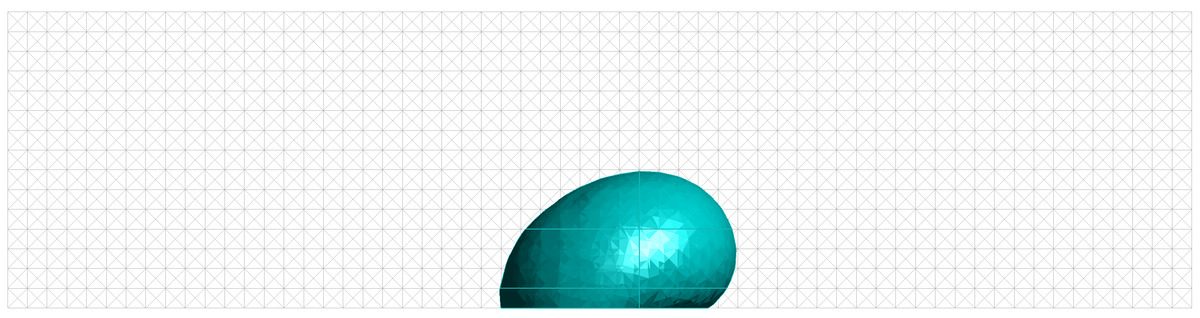}
				\begin{tikzpicture}[overlay,remember picture,scale=0.5]
					\fill[green!80!black] (0,0,0) circle[radius=3.5pt];
					\draw[-{Latex[length=2mm]}, red!80!black, thick] (0,0,0) -- (1,0,0);
					\draw[-{Latex[length=2mm]}, blue!80!black, thick] (0,0,0) -- (0,1,0);
					\node at (1.2,0,0) {\scriptsize$x$};
					\node at (0,1.3,0) {\scriptsize$z$};
					\node at (0.1,0.1,1) {\scriptsize$y$};
				\end{tikzpicture}
			\end{subfigure}
			\begin{subfigure}{\textwidth}           
				\includegraphics[width=\textwidth]{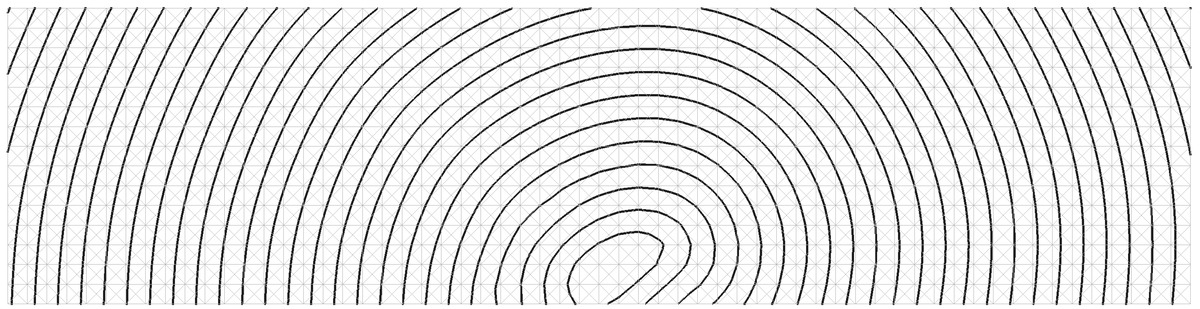}
			\end{subfigure}    
		\end{minipage}
		\hfill
		\begin{minipage}{.25\textwidth}
			\begin{subfigure}{\textwidth}      
				\includegraphics[width=\textwidth]{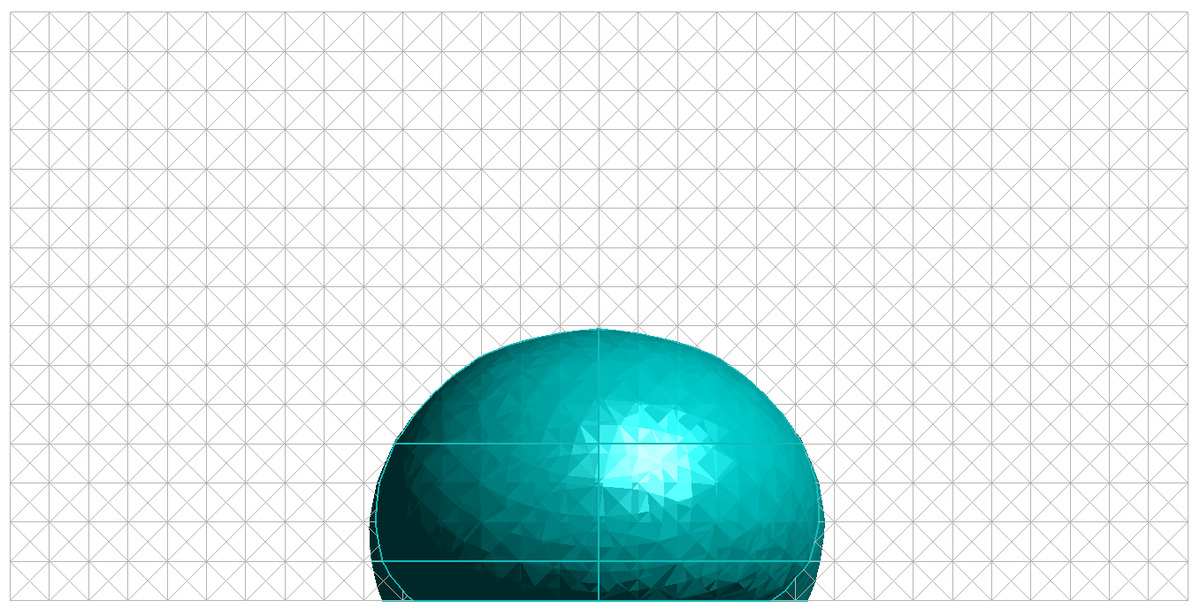}
				\begin{tikzpicture}[overlay,remember picture,scale=0.5]
					\draw[-{Latex[length=2mm]}, green!80!black, thick] (0,0,0) -- (1,0,0);
					\draw[-{Latex[length=2mm]}, blue!80!black, thick] (0,0,0) -- (0,1,0);  
					\fill[red!80!black] (0,0,0) circle[radius=3.5pt];
					\node at (1.2,0,0) {\scriptsize$y$};
					\node at (0,1.3,0) {\scriptsize$z$};
					\node at (0.1,0.1,1) {\scriptsize$x$};
				\end{tikzpicture}
			\end{subfigure}
			\begin{subfigure}{\textwidth}         
				\includegraphics[width=\textwidth]{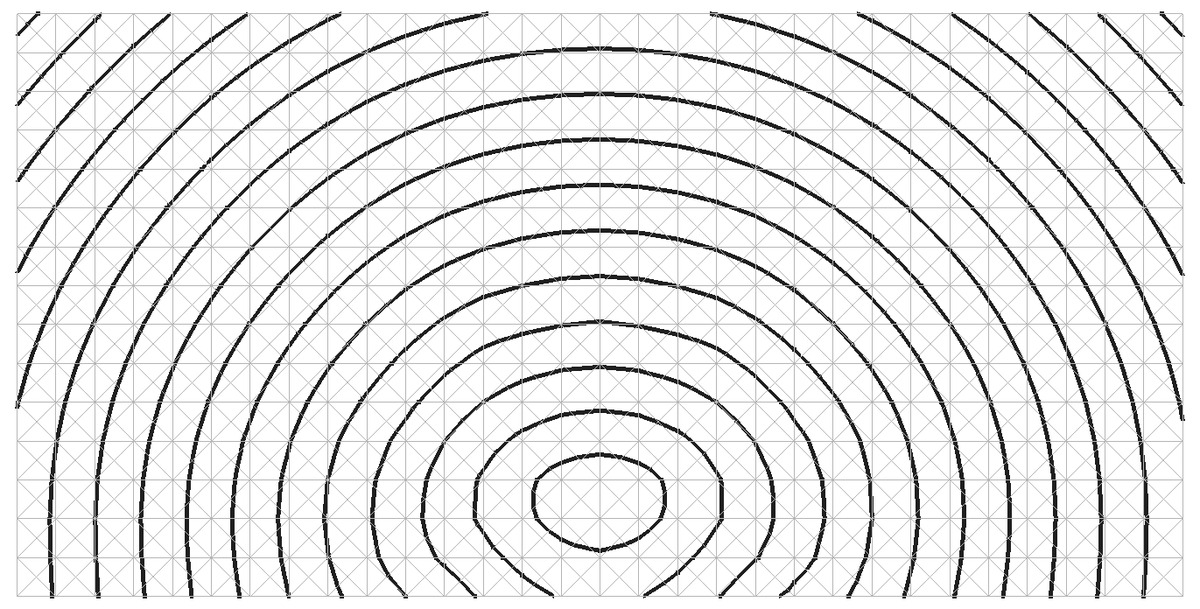}
			\end{subfigure}    
		\end{minipage}
		\caption{Droplet in flow channel. Simulated using the proposed elliptic distance re-initialization.}
		\label{fig:channel_elliptic}
	\end{figure}
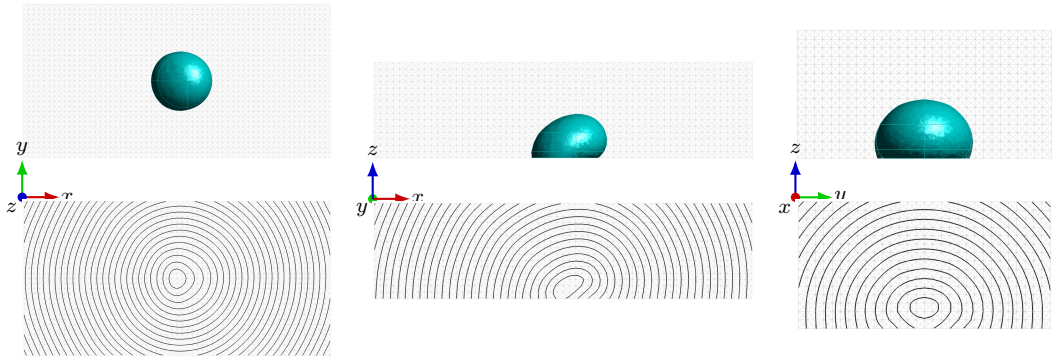

	To further illustrate the effectiveness of our proposed method, Fig.\ref{fig:channel_elliptic} presents iso-phi contours on three distinct plane cross-sections: $xy$, $xz$, and $yz$. Thanks to our approach, which fully reconstructs the level-set, these contours exhibit an exceptional smoothness that allows us to successfully simulate this challenging test case.
	
	It is also worth mentioning that the authors attempted to
	simulate this test case using other level-set re-initialization methods, i.e. the approach proposed by R. N. Elias, M. A. Martins, and A. L. Coutinho~\cite{elias_simple_2007}, as well as the traditional elliptic re-distancing~\cite{basting_minimization-based_2013}. However, these simulations were unsuccessful due to irregularities in the level-set, leading to spurious droplet shape evolution.
	
	\section{Summary and conclusions}
	In this work, a two-step level-set re-initialization technique was presented. In the first step, the level-set function was reconstructed based on a Poisson equation, providing an accurate initial state for the subsequent optimization process. In the second step, the level-set function was regularized using an optimization approach, aided by a well-defined potential. 
	
	It must be noted that, in the presented approach, boundary conditions are needed only for the reconstruction step. Therefore, compared to the often used hyperbolic approaches, the proposed algorithm is less sensitive to the chosen boundary condition, and inaccuracies at the initialization step are corrected by the consecutive elliptic level-set re-initialization.
	
	To ensure accurate re-initialization, a novel geometrical approach was employed to define boundary conditions. Its essential benefit is that it allows to effectively circumvent blind spots, which are critical in maintaining a smooth level-set function, particularly near the contact line.
	
	Additionally, incorporating the element splitting into re-initialization plays crucial role in preserving the location of the zero level-set, which is a fundamental aspect of maintaining interface accuracy.
	
	The proposed method is tested in various scenarios, including static and dynamic droplets, both spherical and ellipsoidal, and droplets confined in channels. The numerical tests carried out confirmed the accuracy and effectiveness of the approach in preserving the level-set function and accurately capturing interface dynamics.
	
	Overall, even though the present work concentrated on the FEM/Level-set formulation, the proposed strategy can be beneficial for other methods that employ level-set for interface tracking.
	
	\section*{Acknowledgment}
	This work was performed within the framework of the DIDRO project (Towards establishing a Digital twin for manufacturing via drop-on-demand inkjet printing, \textit{Proyectos Estratégicos Orientados a la Transición Ecológica y a la Transición Digital} 2022-2024, reference TED2021-130471B-I00 ) supported by the \textit{Agencia Estatal de Investigación, Ministerio de Ciencia, Innovacion e Universidades} of Spain.	
	The authors also acknowledge financial support via the ``Severo Ochoa Programme'' for \textit{Severo Ochoa Centre of Excellence (2019-2023) under the grant CEX2018-000797-S funded by MCIN/AEI/10.13039/501100011033} given to the International Centre for Numerical Methods in Engineering (CIMNE). Pavel Ryzhakov is a Serra Hunter fellow.	
	


\begin{thebibliography}{10}
	\expandafter\ifx\csname url\endcsname\relax
	\def\url#1{\texttt{#1}}\fi
	\expandafter\ifx\csname urlprefix\endcsname\relax\def\urlprefix{URL }\fi
	\expandafter\ifx\csname href\endcsname\relax
	\def\href#1#2{#2} \def\path#1{#1}\fi
	
	\bibitem{mulder_computing_1992}
	W.~Mulder, S.~Osher, J.~A. Sethian, Computing interface motion in compressible
	gas dynamics, Journal of computational physics 100~(2) (1992) 209--228.
	
	\bibitem{osher_fronts_1988}
	S.~Osher, J.~A. Sethian, Fronts propagating with curvature-dependent speed:
	algorithms based on {Hamilton}-{Jacobi} formulations, Journal of
	computational physics 79~(1) (1988) 12--49.
	
	\bibitem{gibou_review_2018}
	F.~Gibou, R.~Fedkiw, S.~Osher, A review of level-set methods and some recent
	applications, Journal of Computational Physics 353 (2018) 82--109.
	
	\bibitem{sussman_level_1994}
	M.~Sussman, P.~Smereka, S.~Osher, A level set approach for computing solutions
	to incompressible two-phase flow, Journal of Computational physics 114~(1)
	(1994) 146--159.
	
	\bibitem{osher_level_2001}
	S.~Osher, R.~P. Fedkiw,
	\href{https://www.sciencedirect.com/science/article/pii/S0021999100966361}{Level
		{Set} {Methods}: {An} {Overview} and {Some} {Recent} {Results}}, Journal of
	Computational Physics 169~(2) (2001) 463--502.
	\newblock \href {https://doi.org/10.1006/jcph.2000.6636}
	{\path{doi:10.1006/jcph.2000.6636}}.
	\newline\urlprefix\url{https://www.sciencedirect.com/science/article/pii/S0021999100966361}
	
	\bibitem{hashemi_toward_2021}
	M.~R. Hashemi, P.~B. Ryzhakov, R.~Rossi, Toward droplet dynamics simulation in
	polymer electrolyte membrane fuel cells: {Three}-dimensional numerical
	modeling of confined water droplets with dynamic contact angle and
	hysteresis, Physics of fluids 33~(12), publisher: AIP Publishing (2021).
	
	\bibitem{noda_extended_2022}
	M.~Noda, Y.~Noguchi, T.~Yamada,
	\href{https://www.sciencedirect.com/science/article/pii/S0045782522000974}{Extended
		level set method: {A} multiphase representation with perfect symmetric
		property, and its application to multi-material topology optimization},
	Computer Methods in Applied Mechanics and Engineering 393 (2022) 114742.
	\newblock \href {https://doi.org/10.1016/j.cma.2022.114742}
	{\path{doi:10.1016/j.cma.2022.114742}}.
	\newline\urlprefix\url{https://www.sciencedirect.com/science/article/pii/S0045782522000974}
	
	\bibitem{hirt_volume_1981}
	C.~Hirt, B.~Nichols,
	\href{https://linkinghub.elsevier.com/retrieve/pii/0021999181901455}{Volume
		of fluid ({VOF}) method for the dynamics of free boundaries}, Journal of
	Computational Physics 39~(1) (1981) 201--225.
	\newblock \href {https://doi.org/10.1016/0021-9991(81)90145-5}
	{\path{doi:10.1016/0021-9991(81)90145-5}}.
	\newline\urlprefix\url{https://linkinghub.elsevier.com/retrieve/pii/0021999181901455}
	
	\bibitem{maric_unstructured_2020}
	T.~Marić, D.~B. Kothe, D.~Bothe,
	\href{https://www.sciencedirect.com/science/article/pii/S0021999120304691}{Unstructured
		un-split geometrical {Volume}-of-{Fluid} methods – {A} review}, Journal of
	Computational Physics 420 (2020) 109695.
	\newblock \href {https://doi.org/10.1016/j.jcp.2020.109695}
	{\path{doi:10.1016/j.jcp.2020.109695}}.
	\newline\urlprefix\url{https://www.sciencedirect.com/science/article/pii/S0021999120304691}
	
	\bibitem{aniszewski_parallel_2021}
	W.~Aniszewski, T.~Arrufat, M.~Crialesi-Esposito, S.~Dabiri, D.~Fuster, Y.~Ling,
	J.~Lu, L.~Malan, S.~Pal, R.~Scardovelli, G.~Tryggvason, P.~Yecko, S.~Zaleski,
	\href{https://www.sciencedirect.com/science/article/pii/S0010465521000175}{{PArallel},
		{Robust}, {Interface} {Simulator} ({PARIS})}, Computer Physics Communications
	263 (2021) 107849.
	\newblock \href {https://doi.org/10.1016/j.cpc.2021.107849}
	{\path{doi:10.1016/j.cpc.2021.107849}}.
	\newline\urlprefix\url{https://www.sciencedirect.com/science/article/pii/S0010465521000175}
	
	\bibitem{hashemi_enriched_2020}
	M.~R. Hashemi, P.~B. Ryzhakov, R.~Rossi,
	\href{http://www.sciencedirect.com/science/article/pii/S004578252030462X}{An
		enriched finite element/level-set method for simulating two-phase
		incompressible fluid flows with surface tension}, Computer Methods in Applied
	Mechanics and Engineering 370 (2020) 113277.
	\newblock \href {https://doi.org/10.1016/j.cma.2020.113277}
	{\path{doi:10.1016/j.cma.2020.113277}}.
	\newline\urlprefix\url{http://www.sciencedirect.com/science/article/pii/S004578252030462X}
	
	\bibitem{maarouf_characteristicsfinite_2022}
	S.~Maarouf, C.~Bernardi, D.~Yakoubi,
	\href{https://www.sciencedirect.com/science/article/pii/S0045782522001530}{Characteristics/finite
		element analysis for two incompressible fluid flows with surface tension
		using level set method}, Computer Methods in Applied Mechanics and
	Engineering 394 (2022) 114843.
	\newblock \href {https://doi.org/10.1016/j.cma.2022.114843}
	{\path{doi:10.1016/j.cma.2022.114843}}.
	\newline\urlprefix\url{https://www.sciencedirect.com/science/article/pii/S0045782522001530}
	
	\bibitem{sussman_coupled_2000}
	M.~Sussman, E.~G. Puckett,
	\href{https://linkinghub.elsevier.com/retrieve/pii/S0021999100965379}{A
		{Coupled} {Level} {Set} and {Volume}-of-{Fluid} {Method} for {Computing} {3D}
		and {Axisymmetric} {Incompressible} {Two}-{Phase} {Flows}}, Journal of
	Computational Physics 162~(2) (2000) 301--337.
	\newblock \href {https://doi.org/10.1006/jcph.2000.6537}
	{\path{doi:10.1006/jcph.2000.6537}}.
	\newline\urlprefix\url{https://linkinghub.elsevier.com/retrieve/pii/S0021999100965379}
	
	\bibitem{solomenko_mass_2017}
	Z.~Solomenko, P.~D.~M. Spelt, L.~Ó~Náraigh, P.~Alix,
	\href{http://www.sciencedirect.com/science/article/pii/S0301932216304669}{Mass
		conservation and reduction of parasitic interfacial waves in level-set
		methods for the numerical simulation of two-phase flows: {A} comparative
		study}, International Journal of Multiphase Flow 95 (2017) 235--256.
	\newblock \href {https://doi.org/10.1016/j.ijmultiphaseflow.2017.06.004}
	{\path{doi:10.1016/j.ijmultiphaseflow.2017.06.004}}.
	\newline\urlprefix\url{http://www.sciencedirect.com/science/article/pii/S0301932216304669}
	
	\bibitem{hashemi_enhanced_2022}
	M.~R. Hashemi, R.~Rossi, P.~B. Ryzhakov,
	\href{https://www.sciencedirect.com/science/article/pii/S0045782522000044}{An
		enhanced non-oscillatory {BFECC} algorithm for finite element solution of
		advective transport problems}, Computer Methods in Applied Mechanics and
	Engineering 391 (2022) 114576.
	\newblock \href {https://doi.org/10.1016/j.cma.2022.114576}
	{\path{doi:10.1016/j.cma.2022.114576}}.
	\newline\urlprefix\url{https://www.sciencedirect.com/science/article/pii/S0045782522000044}
	
	\bibitem{yap_global_2006}
	Y.~F. Yap, J.~C. Chai, T.~N. Wong, K.~C. Toh, H.~Y. Zhang,
	\href{https://doi.org/10.1080/10407790600646958}{A {Global} {Mass}
		{Correction} {Scheme} for the {Level}-{Set} {Method}}, Numerical Heat
	Transfer, Part B: Fundamentals 50~(5) (2006) 455--472.
	\newblock \href {https://doi.org/10.1080/10407790600646958}
	{\path{doi:10.1080/10407790600646958}}.
	\newline\urlprefix\url{https://doi.org/10.1080/10407790600646958}
	
	\bibitem{ge_efficient_2018}
	Z.~Ge, J.-C. Loiseau, O.~Tammisola, L.~Brandt,
	\href{https://www.sciencedirect.com/science/article/pii/S0021999117308136}{An
		efficient mass-preserving interface-correction level set/ghost fluid method
		for droplet suspensions under depletion forces}, Journal of Computational
	Physics 353 (2018) 435--459.
	\newblock \href {https://doi.org/10.1016/j.jcp.2017.10.046}
	{\path{doi:10.1016/j.jcp.2017.10.046}}.
	\newline\urlprefix\url{https://www.sciencedirect.com/science/article/pii/S0021999117308136}
	
	\bibitem{trujillo_distortion_2017}
	M.~F. Trujillo, L.~Anumolu, D.~Ryddner, The distortion of the level set
	gradient under advection, Journal of Computational Physics 334 (2017)
	81--101.
	
	\bibitem{janodet_massively_2022}
	R.~Janodet, C.~Guillamón, V.~Moureau, R.~Mercier, G.~Lartigue, P.~Bénard,
	T.~Ménard, A.~Berlemont,
	\href{https://www.sciencedirect.com/science/article/pii/S0021999122001371}{A
		massively parallel accurate conservative level set algorithm for simulating
		turbulent atomization on adaptive unstructured grids}, Journal of
	Computational Physics 458 (2022) 111075.
	\newblock \href {https://doi.org/10.1016/j.jcp.2022.111075}
	{\path{doi:10.1016/j.jcp.2022.111075}}.
	\newline\urlprefix\url{https://www.sciencedirect.com/science/article/pii/S0021999122001371}
	
	\bibitem{larios-cardenas_deep_2021}
	L.~{\'A}. Larios-C{\'a}rdenas, F.~Gibou,
	\href{https://epubs.siam.org/doi/abs/10.1137/20M1316755}{A {Deep} {Learning}
		{Approach} for the {Computation} of {Curvature} in the {Level}-{Set}
		{Method}}, SIAM Journal on Scientific Computing 43~(3) (2021) A1754--A1779.
	\newblock \href {https://doi.org/10.1137/20M1316755}
	{\path{doi:10.1137/20M1316755}}.
	\newline\urlprefix\url{https://epubs.siam.org/doi/abs/10.1137/20M1316755}
	
	\bibitem{xue_new_2021}
	T.~Xue, W.~Sun, S.~Adriaenssens, Y.~Wei, C.~Liu,
	\href{https://www.sciencedirect.com/science/article/pii/S0021999121002552}{A
		new finite element level set reinitialization method based on the shifted
		boundary method}, Journal of Computational Physics 438 (2021) 110360.
	\newblock \href {https://doi.org/10.1016/j.jcp.2021.110360}
	{\path{doi:10.1016/j.jcp.2021.110360}}.
	\newline\urlprefix\url{https://www.sciencedirect.com/science/article/pii/S0021999121002552}
	
	\bibitem{gaudlitz_improving_2008}
	D.~Gaudlitz, N.~A. Adams,
	\href{https://www.sciencedirect.com/science/article/pii/S0045793007002113}{On
		improving mass-conservation properties of the hybrid particle-level-set
		method}, Computers \& Fluids 37~(10) (2008) 1320--1331.
	\newblock \href {https://doi.org/10.1016/j.compfluid.2007.11.005}
	{\path{doi:10.1016/j.compfluid.2007.11.005}}.
	\newline\urlprefix\url{https://www.sciencedirect.com/science/article/pii/S0045793007002113}
	
	\bibitem{sussman_efficient_1999}
	M.~Sussman, E.~Fatemi, An efficient, interface-preserving level set
	redistancing algorithm and its application to interfacial incompressible
	fluid flow, SIAM Journal on scientific computing 20~(4) (1999) 1165--1191.
	
	\bibitem{zhang_efficient_2020}
	H.~Zhang, L.~Zhao, J.~Mao, X.~Liu,
	\href{https://www.sciencedirect.com/science/article/pii/S0045793020302942}{An
		efficient {3D} iterative interface-correction reinitialization for the level
		set method}, Computers \& Fluids 213 (2020) 104724.
	\newblock \href {https://doi.org/10.1016/j.compfluid.2020.104724}
	{\path{doi:10.1016/j.compfluid.2020.104724}}.
	\newline\urlprefix\url{https://www.sciencedirect.com/science/article/pii/S0045793020302942}
	
	\bibitem{ramanuj_high_2018}
	V.~Ramanuj, R.~Sankaran, High {Order} {Anchoring} and {Reinitialization} of
	{Level} {Set} {Function} for {Simulating} {Interface} {Motion}, Journal of
	Scientific Computing (2018) 1--24.
	
	\bibitem{sethian_fast_2000}
	J.~A. Sethian, A.~Vladimirsky,
	\href{http://www.pnas.org/cgi/doi/10.1073/pnas.090060097}{Fast methods for
		the {Eikonal} and related {Hamilton}- {Jacobi} equations on unstructured
		meshes}, Proceedings of the National Academy of Sciences 97~(11) (2000)
	5699--5703.
	\newblock \href {https://doi.org/10.1073/pnas.090060097}
	{\path{doi:10.1073/pnas.090060097}}.
	\newline\urlprefix\url{http://www.pnas.org/cgi/doi/10.1073/pnas.090060097}
	
	\bibitem{karakus_local_2022}
	A.~Karakus, N.~Chalmers, T.~Warburton,
	\href{https://www.sciencedirect.com/science/article/pii/S0898122122003315}{A
		local discontinuous {Galerkin} level set reinitialization with subcell
		stabilization on unstructured meshes}, Computers \& Mathematics with
	Applications 123 (2022) 160--170.
	\newblock \href {https://doi.org/10.1016/j.camwa.2022.08.010}
	{\path{doi:10.1016/j.camwa.2022.08.010}}.
	\newline\urlprefix\url{https://www.sciencedirect.com/science/article/pii/S0898122122003315}
	
	\bibitem{hysing_new_2006}
	S.~Hysing, \href{http://doi.wiley.com/10.1002/fld.1147}{A new implicit surface
		tension implementation for interfacial flows}, International Journal for
	Numerical Methods in Fluids 51~(6) (2006) 659--672.
	\newblock \href {https://doi.org/10.1002/fld.1147}
	{\path{doi:10.1002/fld.1147}}.
	\newline\urlprefix\url{http://doi.wiley.com/10.1002/fld.1147}
	
	\bibitem{min_reinitializing_2010}
	C.~Min, On reinitializing level set functions, Journal of computational physics
	229~(8) (2010) 2764--2772.
	
	\bibitem{li_level_2005}
	C.~Li, C.~Xu, C.~Gui, M.~D. Fox, Level set evolution without re-initialization:
	a new variational formulation, in: 2005 {IEEE} {Computer} {Society}
	{Conference} on {Computer} {Vision} and {Pattern} {Recognition} ({CVPR}'05),
	Vol.~1, IEEE, 2005, pp. 430--436.
	
	\bibitem{henri_geometrical_2022}
	F.~Henri, M.~Coquerelle, P.~Lubin,
	\href{https://www.sciencedirect.com/science/article/pii/S0021999121005994}{Geometrical
		level set reinitialization using closest point method and kink detection for
		thin filaments, topology changes and two-phase flows}, Journal of
	Computational Physics 448 (2022) 110704.
	\newblock \href {https://doi.org/10.1016/j.jcp.2021.110704}
	{\path{doi:10.1016/j.jcp.2021.110704}}.
	\newline\urlprefix\url{https://www.sciencedirect.com/science/article/pii/S0021999121005994}
	
	\bibitem{della_rocca_level_2014}
	G.~Della~Rocca, G.~Blanquart, Level set reinitialization at a contact line,
	Journal of Computational Physics 265 (2014) 34--49.
	
	\bibitem{elias_simple_2007}
	R.~N. Elias, M.~A. Martins, A.~L. Coutinho, Simple finite element-based
	computation of distance functions in unstructured grids, International
	journal for numerical methods in engineering 72~(9) (2007) 1095--1110.
	
	\bibitem{li_distance_2010}
	C.~Li, C.~Xu, C.~Gui, M.~D. Fox, Distance regularized level set evolution and
	its application to image segmentation, IEEE transactions on image processing
	19~(12) (2010) 3243--3254.
	
	\bibitem{basting_minimization-based_2013}
	C.~Basting, D.~Kuzmin, \href{https://doi.org/10.1007/s00607-012-0259-z}{A
		minimization-based finite element formulation for interface-preserving level
		set reinitialization}, Computing 95~(1) (2013) 13--25.
	\newblock \href {https://doi.org/10.1007/s00607-012-0259-z}
	{\path{doi:10.1007/s00607-012-0259-z}}.
	\newline\urlprefix\url{https://doi.org/10.1007/s00607-012-0259-z}
	
	\bibitem{basting_optimal_2017}
	C.~Basting, D.~Kuzmin, J.~N. Shadid,
	\href{https://onlinelibrary.wiley.com/doi/abs/10.1002/fld.4348}{Optimal
		control for reinitialization in finite element level set methods},
	International Journal for Numerical Methods in Fluids 84~(5) (2017) 292--305.
	\newblock \href {https://doi.org/10.1002/fld.4348}
	{\path{doi:10.1002/fld.4348}}.
	\newline\urlprefix\url{https://onlinelibrary.wiley.com/doi/abs/10.1002/fld.4348}
	
	\bibitem{adams_high-order_2019}
	T.~Adams, S.~Giani, W.~M. Coombs,
	\href{https://www.sciencedirect.com/science/article/pii/S0021999118307915}{A
		high-order elliptic {PDE} based level set reinitialisation method using a
		discontinuous {Galerkin} discretisation}, Journal of Computational Physics
	379 (2019) 373--391.
	\newblock \href {https://doi.org/10.1016/j.jcp.2018.12.003}
	{\path{doi:10.1016/j.jcp.2018.12.003}}.
	\newline\urlprefix\url{https://www.sciencedirect.com/science/article/pii/S0021999118307915}
	
	\bibitem{nitsche_uber_1971}
	J.~Nitsche, \href{https://doi.org/10.1007/BF02995904}{Über ein
		{Variationsprinzip} zur {Lösung} von {Dirichlet}-{Problemen} bei
		{Verwendung} von {Teilräumen}, die keinen {Randbedingungen} unterworfen
		sind}, Abhandlungen aus dem Mathematischen Seminar der Universität Hamburg
	36~(1) (1971) 9--15.
	\newblock \href {https://doi.org/10.1007/BF02995904}
	{\path{doi:10.1007/BF02995904}}.
	\newline\urlprefix\url{https://doi.org/10.1007/BF02995904}
	
	\bibitem{ferrandiz_kratosmultiphysicskratos_2023}
	V.~M. Ferrándiz, P.~Bucher, R.~Zorrilla, R.~Rossi, S.~Warnakulasuriya,
	A.~Cornejo, jcotela, C.~Roig, M.~A. Celigueta, J.~Maria, tteschemacher,
	M.~Masó, G.~Casas, M.~Núñez, P.~Dadvand, S.~Latorre, I.~d. Pouplana, J.~I.
	González, F.~Arrufat, riccardotosi, AFranci, A.~Ghantasala, K.~B. Sautter,
	P.~Wilson, dbaumgaertner, B.~Chandra, A.~Geiser, I.~Lopez, lluís,
	J.~Gárate,
	\href{https://zenodo.org/record/8272092}{{KratosMultiphysics}/{Kratos}:
		{Release} 9.4} (Aug. 2023).
	\newblock \href {https://doi.org/10.5281/zenodo.8272092}
	{\path{doi:10.5281/zenodo.8272092}}.
	\newline\urlprefix\url{https://zenodo.org/record/8272092}
	
	\bibitem{dadvand_object-oriented_2010}
	P.~Dadvand, R.~Rossi, E.~Oñate,
	\href{http://link.springer.com/10.1007/s11831-010-9045-2}{An
		{Object}-oriented {Environment} for {Developing} {Finite} {Element} {Codes}
		for {Multi}-disciplinary {Applications}}, Archives of Computational Methods
	in Engineering 17~(3) (2010) 253--297.
	\newblock \href {https://doi.org/10.1007/s11831-010-9045-2}
	{\path{doi:10.1007/s11831-010-9045-2}}.
	\newline\urlprefix\url{http://link.springer.com/10.1007/s11831-010-9045-2}
	
	\bibitem{hashemi_three_2021}
	M.~R. Hashemi, P.~B. Ryzhakov, R.~Rossi,
	\href{https://www.sciencedirect.com/science/article/pii/S0021999121003752}{Three
		dimensional modeling of liquid droplet spreading on solid surface: {An}
		enriched finite element/level-set approach}, Journal of Computational Physics
	442 (2021) 110480.
	\newblock \href {https://doi.org/10.1016/j.jcp.2021.110480}
	{\path{doi:10.1016/j.jcp.2021.110480}}.
	\newline\urlprefix\url{https://www.sciencedirect.com/science/article/pii/S0021999121003752}
	
	\bibitem{dangla_droplet_2013}
	R.~Dangla, S.~C. Kayi, C.~N. Baroud,
	\href{https://www.pnas.org/doi/full/10.1073/pnas.1209186110}{Droplet
		microfluidics driven by gradients of confinement}, Proceedings of the
	National Academy of Sciences 110~(3) (2013) 853--858, publisher: Proceedings
	of the National Academy of Sciences.
	\newblock \href {https://doi.org/10.1073/pnas.1209186110}
	{\path{doi:10.1073/pnas.1209186110}}.
	\newline\urlprefix\url{https://www.pnas.org/doi/full/10.1073/pnas.1209186110}
	
	\bibitem{liu_microfluidics_2021}
	Y.~Liu, L.~Sun, H.~Zhang, L.~Shang, Y.~Zhao,
	\href{https://doi.org/10.1021/acs.chemrev.0c01289}{Microfluidics for {Drug}
		{Development}: {From} {Synthesis} to {Evaluation}}, Chemical Reviews 121~(13)
	(2021) 7468--7529, publisher: American Chemical Society.
	\newblock \href {https://doi.org/10.1021/acs.chemrev.0c01289}
	{\path{doi:10.1021/acs.chemrev.0c01289}}.
	\newline\urlprefix\url{https://doi.org/10.1021/acs.chemrev.0c01289}
	
	\bibitem{giannitelli_droplet_based_2022}
	S.~M. Giannitelli, E.~Limiti, P.~Mozetic, F.~Pinelli, X.~Han, F.~Abbruzzese,
	F.~Basoli, D.~D. Rio, S.~Scialla, F.~Rossi, M.~Trombetta, L.~Rosanò,
	G.~Gigli, Z.~J. Zhang, E.~Mauri, A.~Rainer,
	\href{https://pubs.rsc.org/en/content/articlelanding/2022/nr/d2nr00827k}{Droplet-based
		microfluidic synthesis of nanogels for controlled drug delivery: tailoring
		nanomaterial properties via pneumatically actuated flow-focusing junction},
	Nanoscale 14~(31) (2022) 11415--11428, publisher: The Royal Society of
	Chemistry.
	\newblock \href {https://doi.org/10.1039/D2NR00827K}
	{\path{doi:10.1039/D2NR00827K}}.
	\newline\urlprefix\url{https://pubs.rsc.org/en/content/articlelanding/2022/nr/d2nr00827k}
	
\end{thebibliography}
	
	
		
		
		

\end{document}